\documentclass{article}
\usepackage{emlines2,bezier}
\usepackage[dvips]{graphicx}
\usepackage{amssymb,amsmath}

\title{Plate-Universe in  Multidimensional  Elastisity Theory}
\author{Sergey S. Kokarev}
\date{Regional Scientific Educational Center "Logos",\\ 150000,
Yaroslavl, Respublikanskaya 80, e-mail: 8helper8@gmail.com}

\begin{document}

\maketitle

\begin{abstract}
 A number of boundary problems in multidimensional
elasticity theory are solved. The solutions can be treated as the
simplest cosmological models. Some specific properties of
the solutions and experimental consequences of the theory are
discussed.
\end{abstract}




\section{Introduction}

In \cite{kok1}-\cite{kok3} an alternative to the concept of space-time dynamics
of general relativity (GR) was developed. It is based on a
unification of the ideas of classical embedding
in a multidimensional space \cite{eiz}-\cite{pav} and elasticity theory
\cite{land1}-\cite{tart}. In the  the framework of this theory the  4-dimensional
space-time is treated as a {\it 4-dimensional elastic plate}, i.e., a
multidimensional body whose sizes along the extra dimensions is much smaller
than along
the four observable ones. It would appear reasonable to treat gravity as
a manifestation of a normal strain of the plate  ({\it bending}), described by
the strain vector components $\xi^m, m=\overline{1,N},$ ($N$ is the number of
extra dimensions), which induces Riemannian geometry on the plate surface. It
can be derived by standard methods of embedding theory \cite{eiz}. On the other
hand, 4-dimensional fields and matter (their energy-momentum tensors) in the
elastic picture are induced by tangent stresses of the plate, which lead to its
{\it stretches} and {\it shears} along the four macroscopic dimensions.

The case of a weak bending, considered in \cite{kok1}, can be related to the
linearized Einstein theory\footnote{ The weakness of bending means physically the validity
of multidimensional Hooke law, or (in GR language) consideration of
gravitational fields far from their sources.}. In this case the mechanical
equilibrium conditions of the plate, obtained by varying the multidimensional free
elastic energy, have the form of {\it inhomogeneous biwave equations}:
\begin{equation}\label{eq}
D_{m}\Box^{2}\xi^{m}=P^{m}, \ \ (m=\overline{1,N}),
\end{equation}
where $D_{m}$ is  the cylindrical stiffness factor of the 4-dimensional plate in
the $m$-th extra dimension, depending on the elastic constants of the theory (see
\cite{kok1}), $P^{m}$ are  the components of an external
multidimensional force normal to the plate surface, bending the plate,
$\Box\equiv\partial_{\mu}\partial^{\mu}$
is the  d'Alembertian in flat Minkowski space-time and there is no summation over $m$
in (\ref{eq}). Together with the Eq.(\ref{eq}), a variational procedure gives the
following boundary conditions:
\begin{equation}\label{gbd}
-D_{m}\oint_{\partial\Gamma}d^{3}S^{\mu}\Box\xi_{m,\mu}\delta\xi^{m}+\frac{D_{m}}{f+1}\oint_{\partial\Gamma}d^{3}S^{\mu}(\xi_{m,\mu,\lambda}+
\eta_{\mu\lambda}f\Box\xi_{m})\delta\xi^{m\ \lambda}_{,}=0,
\end{equation}
where $\partial\Gamma$ is the 3-dimensional boundary of the plate, $f$ is a
dimensionless constant, depending on the elastic properties\footnote{As follows
from \cite{kok1}, $f=\sigma/(1-3\sigma)$.} (the Poisson coefficient $\sigma$) of
the plate medium, $\eta_{\mu\lambda}$ is the original Minkowski metric of
the unstrained plate $M_{1,3}$ (see fig.\ref{plate}).

When the plate Poisson coefficient is $\sigma=1/2$, the free energy
functional transforms into the Einstein gravitational action $S_{g}$ of linearized
GR \cite{land2}, which is in turn exactly the surface term with respect to
the embedding variables $\xi^m$. Dimensional analysis gives a number of possible
relations between the multidimensional parameters of the plate, the Young modulus
$E$ and the thicknesses $\left\{h_m\right\}$,  and the 4-dimensional fundamental
constants. The most realistic relation, obtained in \cite{kok1}, $1/\varkappa\sim
Eh^{N+3}$, where $h$ is the average thickness of the plate and $\varkappa$ is the Einstein
gravitational constant, supports  Sakharov's idea on a relation between
$\varkappa$ and elastic properties of space-time \cite{sach}.

The aim of the present article is an analysis of distinctive properties of the
simplest space-time models, based on Eqs.(\ref{eq}), (\ref{gbd}).

In Sec.\ref{model} the model to be considered is described.

In Sec.\ref{bcond} we describe the boundary conditions,
their notations and classes.

In Sec.\ref{plane} the  case of an unbounded plate is considered.

In Sec.\ref{hplane},\ref{quad},\ref{band},\ref{sbrec} we consider half-plane,
quadrant, band, semi-band and rectangle plate, respectively.

In Sec.\ref{wave} the case of a wave solution is analyzed.

In Sec.\ref{concl} some observable consequences of the models are discussed.

\section{The model}\label{model}

\begin{figure}[htb]
\centering
\input{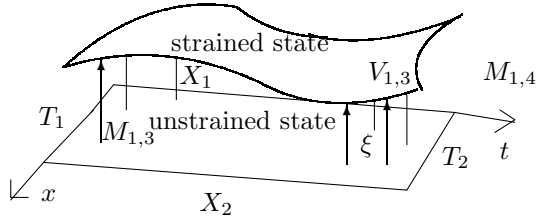}
\caption{\small Factorized surface of a 4-dimensional plate. The plate in an unstrained
state (at the bottom) with pseudo-Euclidean intrinsic geometry, is
Minkowski space-time $M_{1,3}$. The plate in a strained state (at the top)
with Riemannian intrinsic geometry, is a  curved space-time
$V_{1,3}$. $T_1,T_2$ are  possible time-like boundaries, $X_1, X_2$ are space-like
ones. Arrows show the strain vector field defined on $M_{1,3}$.} \label{plate}
\end{figure}

Let us consider pseudo-Euclidean space $M_{1,4}$ of five dimensions with  one
time-like and four space-like ones. Its metric can be written as
\[\eta_{AB}=\text{diag}\{+1,-1,-1,-1,-1\}.\] Flat 4-dimensional space-time can be
realized as the Min\-kow\-ski plane $M_{1,3}$, embedded in $M_{1,4}$, so, that its
equation is
\begin{equation}\label{pleq}
x^{5}=0
\end{equation}
and its induced 4-dimensional metric
$$
 \eta_{\mu\nu}=\mbox{\rm diag}\{+1,-1,-1,-1\}
$$
(Fig.\ref{plate}). We consider $M_{1,3}$ as a middle plane of a physical
4-dimensional thin plate in an unstrained state.

Let $\overrightarrow{\xi}$ be a smooth vector field defined on $M_{1,3}$, orthogonal to
$M_{1,3}$. This vector field determines a deformation of $M_{1,3}$ into some
curved Riemannian manifold $V_{1,3}$ (up to a parallel transition and rigid
rotation in $M_{1,4}$)\footnote{Anywhere below  we implicitly ommit such
unessential rigid transitions and rotations of the plate in embedding space.}.
The only nonzero component of $\overrightarrow{\xi}$ is $\xi^{5}\equiv\xi$ (due to
(\ref{pleq}) and orthogonality to $M_{1,3}$). We postulate its special
dependence on the 4-dimensional coordinates $\xi=\xi(t,x)$.\footnote{The dependence
$\xi=\xi(t)$, which is simpler, leads to a cylindrical bending of the
plate and pseudo-Euclidean inner geometry. However, physically, a cylindrically
bent plate is not equivalent to a flat unstrained one, in spite of the identity
of their inner geometries.} Under straining of this kind it is convenient
to factorize all points of the plate along the $y$ and $z$ directions. Such a factorized
strained 4-dimensional plate can be considered as a 2-dimensional surface with
the coordinate map $(t,x)$ (Fig.\ref{plate}). There is an evident analogy of  such
a plate with a moving classical string \cite{green}. In contrast to the latter,
the dynamical equations of the plate (\ref{eq}) are derived from minimality of
free elastic energy rather then minimal surface condition.

To avoid metaphysical problem of the nature of  external multidimensional forces,
 we put $P_m=0$, i.e., the only homogeneous biwave equation
\begin{equation}\label{heq}
\Box^{2}\xi=0
\end{equation}
will be considered. Throughout this paper we assume,
that stiffness factor is $D\neq 0\
(f\neq-1)$.

Due to homogeneity of both (\ref{heq}) and the boundary conditions (\ref{gbd}), all
solutions will contain an arbitrary multiplicative constant, which is implicitly
put equal to 1 everywhere below\footnote{The dimensional amplitude of the strain vector
components $\xi^m$ must be much smaller then the corresponding thickness $h_m$ of the
plate to satisfy the bending weakness requirements. Thus all graphs of $\xi$
presented in the paper are given in an enlarged scale.}.

In this paper we consider only the case of rectangle
symmetry, which admits a surface unbounded or bounded in one or more
directions, with parallel or perpendicular (in pseudo-Euclidean
sense) boundaries.
Such a surface can be treated, e.g., as an elastic
model of the whole Universe or its region. Cosmological
consequences of the model will be considered in section 12.

\section{Boundary conditions}\label{bcond}

Let us designate the  type of boundary conditions of the model by parethesis with four
symbols, which characterize this type at the boundaries $T_1,X_1,T_2,$ $X_2$
respectively (fig.\ref{plate}). Thus, the case of an unbounded plate
will be labelled  $(\infty,\infty,\infty,\infty)$. The case with
one boundary $T_1$ will be designated
$(\Gamma,\infty,\infty,\infty)$ etc. Any symbol can be $\infty$ or $\Gamma$, so
we have $2^4=16$ plates types. This types, by the symmetry between $x$ and $t$,
are combined in the following six classes:

$1. (\infty,\infty,\infty,\infty);$

$2. (\Gamma,\infty,\infty,\infty), (\infty,\Gamma,\infty,\infty),$

\hspace{10pt}$(\infty,\infty,\Gamma,\infty), (\infty,\infty,\infty,\Gamma);$

$3. (\Gamma,\Gamma,\infty,\infty), (\infty,\infty,\Gamma,\Gamma),$

\hspace{10pt}$(\infty,\Gamma,\Gamma,\infty), (\Gamma,\infty,\infty,\Gamma),$

$4. (\Gamma,\infty,\Gamma,\infty), (\infty,\Gamma,\infty,\Gamma).$

$5. (\infty,\Gamma,\Gamma,\Gamma), (\Gamma,\infty,\Gamma,\Gamma),$

\hspace{10pt}$(\Gamma,\Gamma,\infty,\Gamma), (\Gamma,\Gamma,\Gamma,\infty).$

$6. (\Gamma,\Gamma,\Gamma,\Gamma).$

In this paper we shall use three types of boundary conditions: a {\it
pinned boundary} ($P$), a {\it free supported boundary} ($S$) and a {\it free boundary}
($F$) by analogy with standard elasticity theory \cite{land1}. These
conditions are expressed by the following requirements:
$\xi|_{\Gamma}=\partial_{\vec n}\xi|_{\Gamma}=0$ ($P$), where $\partial_{\vec
n}$ is derivative normal to the boundary; the equality $\xi|_{\Gamma}=0$ and
vanishing of the second integrand in (\ref{gbd}) ($S$); vanishing of both
integrands in (\ref{gbd}) ($F$). So the symbol $\Gamma$  can take three
value $F,S,P$. The second class gives 3 different boundary condition, the third one ---
6 (in view of the symmetry between $x$ and $t$), the fourth one --- 6, the fifth one --- 18,
the sixth one
--- 21 and altogether 55 different boundary-value  problems. Due to incompatibility
of certain boundary conditions,  a considerable part of them gives only trivial solutions.

For our model Eq.(\ref{gbd}) can be simplified, and the three
boundary conditions under consideration can be put in the form:

\[(P)\ \ \xi=0|_{\Gamma}\ \  \mbox{and}\]

\[\left\{\begin{array}{r}
                                \xi,_t|_{\Gamma}=0;\\
                                \xi,_x|_{\Gamma}=0.
                                \end{array}
                                \right.\]

\[(S)\ \ \xi|_{\Gamma}=0\  \ \mbox{and}\]

\begin{equation}\label{bp}
\left\{\begin{array}{r} (1+f)\xi,_{tt}-f\xi,_{xx}=0|_{\Gamma};\\
(1+f)\xi,_{xx}-f\xi,_{tt}=0|_{\Gamma}.
\end{array}
\right.
\end{equation}

\[(F)\  \ \xi,_{xt}=0|_{\Gamma}\ \ \mbox{and}\]
\[
\left\{
\begin{array}{rcr}
\Box\xi,_t|_{\Gamma}=0&\mbox{and}&(1+f)\xi,_{tt}-f\xi,_{xx}=0;|_{\Gamma}\\
\Box\xi,_x|_{\Gamma}=0&\mbox{and}&(1+f)\xi,_{xx}-f\xi,_{tt}=0|_{\Gamma},
\end{array}
\right.\]

where the upper lines in brackets are related to $T$-boundaries, and the lower ones to
$X$-boundaries.

\section{Class 1: boundless plate}\label{plane}

Let us start from the simplest case of a plate infinite
in all four dimensions. Such a plate
in the unstrained state can be imagined
as a boundless plane. It corresponds to a model
of a universe  infinite in space and time.
The equilibrium equation (\ref{heq})
of the plate for the strain vector
$\xi(t,x)$
takes the following form:
\begin{equation}\label{dal}
\Box^{2}\xi=(\partial_{t}^{2}-\partial_{x}^{2})^{2}\xi=0.
\end{equation}
In the new variables $u=x+t,\ v=x-t$,
Eq. (\ref{dal})
takes the canonic form:
\begin{equation}\label{dalc}
\partial_{u}^{2}\partial_{v}^{2}\xi=0.
\end{equation}
Integrating (\ref{dalc}) consequtively over  $u$ and $v$,
we get the general solution:
\begin{equation}\label{gsol}
\xi=uV(v)+vU(u)+M(u)+L(v),
\end{equation}
where  $V,U,M,L$ are arbitrary functions, which can be founded by
specifying the four initial (boundary) conditions. For arbitrary
initial conditions:
\begin{equation}\label{begc}
\xi(x,0)=f_{0}(x),\ \ \xi_{t}(x,0)=f_{1}(x),\ \
\xi_{tt}(x,0)=f_{2}(x),\ \ \xi_{ttt}(x,0)=f_{3}(x)
\end{equation}
the general solution takes the form:
\[
V=\frac{1}{8}(f_{0}'-f_{1}-I[f_{2}]+I^{2}[f_{3}]);\]

\[U=\frac{1}{8}(f_{0}'+f_{1}-I[f_{2}]-I^{2}[f_{3}]);\]

\[M=\frac{1}{2}\left(f_{0}+\frac{3}{2}I[f_{1}]-\frac{1}{2}I^{3}[f_{3}]+\right.\]

\begin{equation}\label{sys}
\left.\frac{x}{4} (-f_{0}'-f_{1}+I[f_{2}]+I^{2}[f_{3}])\right);
\end{equation}

\[
L=\frac{1}{2}\left(f_{0}-\frac{3}{2}I[f_{1}]+\frac{1}{2}I^{3}[f_{3}]+\right.\]

\[\left.\frac{x}{4}
(-f_{0}'+f_{1}+I[f_{2}]-I^{2}[f_{3}])\right),\] where the prime
denotes differentiation of a function in its argument and
\[I^{n}[f_{i}]=\underbrace{\int dx\dots\int dx}_{n}f_{i}(x)\]
is the $n$-th antiderivative of the function $f_{i}$. Eqs. (\ref{gsol})
and (\ref{sys}) give a generalization of the well-known d'Alembert formula
for unbounded string oscillation \cite{mmf}.

From the form of  (\ref{gsol}) it follows
that solutions of the biwave equation in the presence
of  nonzero functions $V$ and $U$
do not describe the propagation of an initial disturbance
as a wave with invariable front.

Let us derive the internal geometry of the resulting  curved  surface of the plate.
Substituting $\xi$ from Eq. (\ref{gsol})
into the expression for themetric in embedding theory
(see \cite{kok1}):
\begin{equation}\label{metr}
ds^{2}=\eta_{\mu\nu}dx^{\mu}dx^{\nu}+
\eta_{mn}\xi^{m}_{,\mu}\xi^{n}_{,\nu}dx^{\mu}dx^{\nu},
\end{equation}
the line element on the surface of the strained 2D (factorized) plate
after some transformations can be written in the following form: \footnote{When
the embedding space has N extra dimensions, the induced metric (\ref{metrpl}) takes the
following more general form: \[
ds^{2}=\overrightarrow{\Upsilon}^{2}du^{2}+\overrightarrow{\Theta}^{2}dv^{2}-(1+2
\overrightarrow{\Upsilon}\cdot\overrightarrow{\Theta} )du\,dv, \] where
\[
\overrightarrow{\Upsilon}=\{\Upsilon^{m}\},\
\Upsilon^{m}=V^{m}(v)+{vU^{m}}'(u)+{M^{m}}'(u);\] \[
\overrightarrow{\Theta}=\{\Theta^{m}\},\ \Theta^{m}=U^{m}(u)+u{V^{m}}'(v)+{L^{m}}'(v),
\] $m=\overline{1,N}$ and the scalar product is calculated in flat metric
$\eta_{mn}$ of the subspace orthogonal to the original Minkowski plane
$M_{1,3}$ (see \cite{kok1}). The curvature will take the form
\begin{equation}\label{curv}
R_{0101}=\eta_{mn}(\xi_{,0,0}^{m}\xi_{,1,1}^{n}-\xi^{m}_{,0,1}\xi^n_{,0,1})
\end{equation}
}
\begin{equation}\label{metrpl}
ds^{2}=-(\Upsilon^{2}du^{2}+\Theta^{2}dv^{2})+(2\Upsilon\cdot \Theta-1)dudv,
\end{equation}
where
\[
\Upsilon=V(v)+vU'(u)+M'(u);\ \ \Theta=V(u)+uV'(v)+L'(v),
\]
the prime denotes differentiation a function with respect to its
argument.
The nonzero component of the linearized Riemannian tensor is
\begin{equation}\label{curvpl}
{}^4R_{0101}=-(\xi_{,0,0}\xi_{,1,1}-(\xi_{,0,1})^{2}).
\end{equation}
Using 4-dimensional Minkowski metric one can verify that
the  Einstein conservative tensor vanishes,
so the  solution (\ref{metrpl})
corresponds to vacuum solutions of the linearized
Einstein theory.

Let us take the simplest initial conditions:
\[
\xi(x,0)=f(x);\]
\[
\xi_{t}(x,0)=\xi_{tt}(x,0)=\xi_{ttt}(x,0)=0.
\]
the corresponding solution has the form: \[
\xi=\frac{1}{2}[f(x+t)+f(x-t)]+\]
\begin{equation}\label{gdal}
\frac{t}{4}[f'(x-t)-f'(x+t)].
\end{equation}
Let, for example, $f=\sin x$. Then $\xi$ can be reduced to the form:
\begin{equation}\label{eiiii}
\xi=\sin x(\cos t+\frac{t}{2}\sin t)
\end{equation}
The solution is shown in Fig.\ref{fiiii}.

\section{Class 2: semibounded plate}\label{hplane}

Consider a model, beginning at the time $t=0$ but having no  end
(or vice versa). In an unstrained state such a plate can be presented by a half-plane,
with the edge $t=0.$ To find a solution, it is necessary to take into account the boundary
conditions (\ref{bp}) on the hyphersurface  $t=0$.

1) The case $(P,\infty,\infty,\infty)$. The boundary conditions (\ref{bp}) (P) mean
that $f_{0}$ and $f_{1}$ in (\ref{gsol})-(\ref{sys}) should be put equal to zero.

The plate with $(P,\infty,\infty,\infty)$ boundary under $f_{2}=\sin x,\ \ f_{3}=0$,
\begin{equation}\label{epiii}
\xi=t\sin t\sin x
\end{equation}
is shown in Fig.\ref{fpiii}.

2) The case $(S,\infty,\infty,\infty)$. The conditions (\ref{bp}) (S) give in
(\ref{begc}): $f_{0}=0, \ f_{2}=0.$ For $f_1=\sin x,\ f_{2}=f_3=f_4=0$ we get
\begin{equation}\label{esiii}
\xi=\sin x(3\sin t-t\cos t).
\end{equation}

3)  The case $(F,\infty,\infty,\infty)$. The conditions (\ref{bp}) (F) take the
form:
\[ -xV''+xU''+M''-L''=0;\]
\begin{equation}\label{gfiii}
V''-U''=0;
\end{equation}
\[xV''+xU''-p(V'+U')+M''+L''=0,\]
which leads to the general solution
\begin{equation}\label{gsolfiii}
V=U;\ \ M=L;\ \ L=(p+2)I[U]-xU,
\end{equation}
depending on one arbitrary function $U$. In (\ref{gfiii}) and below
$p\equiv2(1+2f)$. For $U=\sin x$,
\[ V=\sin x,\ \ L=M=-4(1+f)\cos x-x\sin x \]
and
\begin{equation}\label{efiii}
\xi=\cos x(t\sin t+2(1+f)\cos t).
\end{equation}

\section{Class 3: semibounded in time and space plate}\label{quad}

In the case of two edges orthogonal to each other, $t=0$ and $x=0,$
we have a plate-quadrant. Such a plate can be considered as a
model of universe, having a  beginning in space and time directions.
To obtain solutions, we shall use the general expression
(\ref{gsol}).
Let us consider six possible boundary-value problems.

1) The case $(P,P,\infty,\infty)$. Eq. (\ref{gsol}) with conditions (\ref{bp})
(P) give the following system:

\[(1)\ xV+xU+M+L=0;\]
\begin{equation}\label{eqppii}
(2)\ V-U+x(U'-V')+M'-L'=0;
\end{equation}
\[(3)\ x\bar V-xU+M+\bar L=0;\]
\[(4)\ \bar V+U+x(\bar V'-U')+M'+\bar L'=0;\]
here and below $\bar
f(x)=f(-x)$. It is easy to show that (\ref{eqppii}) has only
a trivial solution: $\xi=0$.

2) The case $(S,S,\infty,\infty)$. The conditions (\ref{bp}) (S) lead to the
system:

\[(1)\ xV+xU+M+L=0;\]
\[(2)\ x(U''+V'')-p(V'+U')+M''+L''=0;\]
\begin{equation}\label{eqssii}
(3)\ x\bar V-xU+M+\bar L=0;
\end{equation}
\[(4)\ x(\bar V''-U'')+p(\bar V'+U')+M''+\bar L''=0;.\]
Its general solution is \[ V=A=-U;\ \ L=S=-M;\] here and below $A(x),\
S(x)$
are arbitrary odd and even functions, respectively. The strain vector $\xi$ can
be put in the form:
\[\xi=S(x-t)-S(x+t)+x(A(x-t)-\]
\begin{equation}\label{gssii}
A(x+t))+t(A(x-t)+A(x+t)).
\end{equation}
For $S=\cos x$, $A=\sin x,$ we get
\begin{equation}\label{essii}
 \xi=\sin x(\sin t+t\cos t)-x\cos x\sin t,
\end{equation}
which is shown in
Fig.\ref{fssii}.

3) The case $(F,F,\infty,\infty)$. The conditions (\ref{bp}) gives the
following system:

\[(1)\ U''-V''=0;\]
\[(2)\ x(U''+V'')-p(V'+U')+M''+L''=0;\]
\begin{equation}\label{eqffii}
(3)\ x(U''-V'')+M''-L''=0;
\end{equation}
\[(4)\  \bar V''+U''=0;\]
\[(5)\ x(\bar V''-U'')+p(\bar V'+U')+M''+\bar L''=0;\]
\[(6)\ -x(\bar V''+U'')+M''-\bar L''=0.\]

The general solution is \[ V=U=x^{p+1};\ \ \ L=M=0\] and the strain vector
\begin{equation}\label{gffii}
\xi=(x^2-t^2)((x-t)^p+(x+t)^p).
\end{equation}
Here \[ p+1=\frac{2m+1}{2n+1},\   m,n\in \mathbb{Z}. \] The case
$m=2,\ n=0$:
\begin{equation}\label{effii}
\xi=(x^2-t^2)(x^4+t^4+6x^2t^2)
\end{equation}
is plotted in Fig.\ref{fffii}.

4) The case $(P,S,\infty,\infty)$. the boundary conditions are given by
Eqs. (1), (2) from (\ref{eqppii}) and Eqs. (3), (4) from
(\ref{eqssii}). The general solution is
\begin{equation}\label{scond}
V=A=-U;\ L=S=-M;\ \  S=2I[A]-xA,
\end{equation}
and the strain vector has the same form as in (\ref{gssii}), with the supplementary
condition (\ref{scond}) between $S$ and $A$. The case $A=\sin x$ and $S=-2\cos
x-x\sin x,$ with
\begin{equation}\label{epsii}
\xi=\sin x(t\cos t-\sin t)
\end{equation}
is shown in Fig.\ref{fpsii}.

5) The case $(P,F,\infty,\infty)$. The boundary conditions are given by Eqs.
(1), (2) from (\ref{eqppii}) and Eqs. (4)-(6) from (\ref{eqffii}). The general
solution is
\[ V=x^\delta;\ U=-\varepsilon x^\delta;\ \   M=\Delta
x^{\delta+1};\ L=-\varepsilon\Delta x^{\delta+1},\ \] where \[
\delta=\frac{\varepsilon+p+1}{2}=\frac{4k+1-\varepsilon}{2(2l+1)};\ k, l\in\
\mathbb{Z};\ \ \Delta=\frac{-1+\varepsilon(1+p)}{\varepsilon+p+3};\ \
\varepsilon=\pm1.\] The strain vector has the form:
\begin{equation}\label{gpfii}
\xi=(x-t)^\delta(x(1+\varepsilon)+t(p+2))-\varepsilon(x+t)^\delta(x(1+\varepsilon)-t(p+2)).
\end{equation}
The case $k=1,\ l=0,\ \varepsilon=-1$,
\begin{equation}\label{epfii}
\xi=t^2(3x^2+t^2)
\end{equation}
is shown in Fig.\ref{fpfii}.

6) The case $(F,S,\infty,\infty)$. The boundary conditions are given by Eqs.
(1)-(3) from (\ref{eqffii}) and Eqs. (3), (4) from (\ref{eqssii}). The general
solution is:
\begin{equation}\label{gfsii}
U=V=S;\ \ M=L=A=-xS-2I[S],
\end{equation}
and $f=-3/2$.

\section{Class 4: infinite bandlike plate}\label{band}

Consider a plate with a beginning at $t=0$ and an end at $t=\tau$ in time,
but spatially infinite.
In this case both boundary
conditions should be formulated.
In the problem with two opposite bounds, an analogue of the d'Alambert formulae
(\ref{gsol}) is invalid, because of effects of repeated reflections from
the boundaries.
We shall solve the biwave equation
by separating variables.

Let us find a solution of biwave equation in the form of a series:
\begin{equation}\label{row}
\xi(x,t)=\sum\limits_{k=0}^\infty T_{k}(t)\cdot R_{k}(x),
\end{equation}
where $\{T_k\}$ and $\{R_k\}$ is some  complete set of linearly independent functions on
 $[0;\tau]\times\mathbb{R}$
\cite{kolm}. Substituting (\ref{row})
into the biwave equation we obtaine for any $k$:
\[
\stackrel{\cdots\cdot}{T}R-2\Delta R\ddot T+T\Delta^{2}R=0
\]
or, dividing by $RT$
\begin{equation}\label{sep}
\frac{\stackrel{\cdots\cdot}{T}}{T}-2\frac{\Delta R}{R}\frac{\ddot T}{T}+\frac{\Delta^{2}R}{R}=0
\end{equation}
This equation can be satisfied in two ways\footnote{Uniquess of the
solutions obtained is not guaranteed by the method.}:

A)
\[
\ddot T=\lambda T;\ \ (\Delta-\lambda)^{2}R=0.
\]

B)
\[
\Delta R=\lambda R;\ \ \stackrel{\cdots\cdot}{T}-2\lambda \ddot T +\lambda^{2}=0,
\]
where $\lambda$ is a separation constant.

In our case of timelike bounds it is necessary to use the
case (B) to satisfy all boundary conditions.
The general solution of the time-part equations in case (B)
has the form:
\begin{equation}\label{tsol}
T=(\alpha_{1}t+\beta_{1})e^{\sqrt{\lambda}t}+(\alpha_{2}t+\beta_{2})e^{-\sqrt{\lambda}t},
\end{equation}
where  $\alpha_{1},\beta_{1},\alpha_{2},\beta_{2}$ are yet unspecified  complex
integration constants. They will be found after boundary conditions
imposing. The Spatial wave-like equation of $(B)$ has the standard solution
\begin{equation}\label{ssol}
R(x)=Ae^{i\sqrt{\lambda}x}+Be^{-i\sqrt{\lambda}x}=a\cos\sqrt{\lambda}x+b\sin\sqrt{\lambda}x.
\end{equation}

Let us consider all 6 possible boundary conditions of Class 4.

1) The case $(P,\infty,P,\infty)$.
The conditions (\ref{bp}) give
\begin{equation}\label{bpipi}
T(0)=T(\tau)=\dot T(0)=\dot T(\tau)=0.
\end{equation}
It easy to  show that the linear set of equation for the integration constant in
(\ref{tsol}) has only a trivial solution, and so $T\equiv0$.

2) The case $(S,\infty,S,\infty)$. Boundary conditions (\ref{bp}) (S) take the
form
\begin{equation}\label{bsisi}
T(0)=T(\tau)=\ddot T(0)=\ddot T(\tau)=0
\end{equation}
which lead to the following
time-part solution:
\begin{equation}\label{tsisi}
T_k=\sin\omega_{k}t;\ \ \omega_{k}=\frac{\pi k}{\tau},\ k\in\mathbb{Z}
\end{equation}
The constants $a_k, b_k$  in  Eqs. (\ref{row})and (\ref{ssol}) should be
determined by the profile of the plate at some intermediate instant $0<t<\tau$.
This profile must be a periodic function of coordinate $x$ with the main frequency
$\omega_1=\pi/\tau$. An example with $k=1,\ \tau=\pi,\ a=b=1$:
\begin{equation}\label{efifi}
\xi=\sin t(\cos x+\sin x)
\end{equation}
is shown in Fig.\ref{fsisi}.

3)  The case $(F,\infty,F,\infty).$ The boundary conditions (\ref{bp}) (F) yield
the relations
\begin{equation}\label{bfifi}
(f+1)\ddot T-\lambda fT=0;\ \ \ \stackrel{\cdots}{T}=\dot T=0 \ \ \mbox{on} \ \
\partial\Gamma,
\end{equation}
which give only a trivial solution.

4) The case $(P,\infty,S,\infty)$.
Such conditions are provided by (\ref{bpipi})
at $t=0$ and (\ref{bsisi}) at $t=\tau$:
\begin{equation}\label{bpisi}
T(0)=\dot T(0)=T(\tau)=\ddot T(\tau)=0,
\end{equation}
which give only a trivial solution.

5) The case $(P,\infty,F,\infty)$. A combination of (\ref{bpipi}) and
(\ref{bsisi}) gives the following expressions:
\[T(0)=\dot T(0)=0;\]
\begin{equation}\label{bpifi}
(f+1)\ddot T(\tau)-\lambda fT(\tau)=0;
\end{equation}
\[\stackrel{\cdots}{T}(\tau)=\dot T(\tau)=0,\]
which give only a trivial solution.

6) The case $(S,\infty,F,\infty).$ A combination of the cases
(\ref{bsisi}) and (\ref{bfifi}) gives the following system:
\[T(0)=\ddot T(0)=0;\]
\begin{equation}\label{bsifi}
(f+1)\ddot T(\tau)-\lambda fT(\tau)=0;
\end{equation}
\[\stackrel{\cdots}{T}(\tau)=\dot T(\tau)=0,\]
which again gives only
a trivial solution.

Note, that all the results remain valid after the change $t\leftrightarrow x$.

\section{Class 5 and 6: semi-band and rectangular plates.}\label{sbrec}

In this section,  plates with three --- $(\Gamma,\Gamma,\Gamma,\infty)$  and
four --- $(\Gamma,\Gamma,\Gamma,\Gamma)$ boundaries are considered. In the
following investigations we shall use results of the previous section. So,
the $x$-part of any solution for the plate  with boundaries
$(\Gamma,\infty,\Gamma,\infty)$ is presented by the set of functions
(\ref{ssol}). A simple analysis shows that only nonzero cases are
$(\Gamma,S,\Gamma,\infty)$ ($a=0$ in (\ref{ssol})) and $(\Gamma,S,\Gamma,S)$
($\omega=\pi m/l$, $l$ is the  spatial size of the plate). From the previous
paragraph it follows that there are only two types of a boundaries which
yield nontrivial solutions: $(S,S,S,\infty)$ and $(S,S,S,S)$
correspondingly. They are  described by a single relation:
\begin{equation}\label{essss}
\xi=\sin\omega t\sin\omega x
\end{equation}
with the condition $\omega=\pi n/\tau$ in both cases and the additional condition
$\omega=\pi m/l$ for the  case of rectangular plate. In this case a nontrivial
deformation picture exists only if timelike and spacelike lengths of the plate
are in a rational proportion: $l/\tau=m/n,\ m,n\in\mathbb{N}$.

An examples with $l=\pi,\ \tau=\pi,\ m=n=1$ are shown in Fig.
\ref{fssss}.

\section{Wave-like solutions.}\label{wave}

As has been mentioned above,  solutions of the homogeneous biwave
equation possess more generality than those of the wave one.
From (\ref{gsol}) it is easy  to notice that
the wave solution of biwave equation corresponds to a special
choice of the initial conditions, i.e., to a special
form of the functions $f_{i}$.
Let us determine those  initial conditions
which give common wave properties to the solutions of
the biwave equations.

For this purposes put $U=V=0$. Then from the first two expressions
in
(\ref{sys}) it follows:
\begin{equation}\label{cond}
f_{0}''=f_{2};\ \ f_{1}''=f_{3},
\end{equation}
or, in invariant form, $\Box\xi|_{t=0}=0,\ \ \Box\xi_{,t}|_{t=0}=0$. For the
remained functions $M$ and $L$ we get from (\ref{sys}): $$
M=\frac{1}{2}(f_{0}+I[f_{1}]);\ L=\frac{1}{2}(f_{0}-I[f_{1}]) $$ that is
the standard d'Alambert formula.

The wave interpretation of the  biwave equation can be made clear
in the following way. Put $\varphi=\Box\xi$, then the biwave equation
for $\xi$ takes the form of the wave equation for $\varphi$:
$\Box\varphi=0$. If $\varphi=j$ is its solution satisfying
all boundary conditions, then the biwave equation takes the form:
$\Box\xi=j$, where $j$  is an effective source, which appears from
the high derivatives of the biwave equation. Under the condition (\ref{cond})
this effective source vanishes. Since $j$ is a solution
of the free wave equation, the source of $\xi$ is a wave.
In other words, the strain vector, satisfying the homogeneous biwave
equation is a wave, created  by another wave.

\section{Observable effects and experimental
verification}\label{concl}

The previous investigation shows that even  the simplest model of the theory gives
a great variety of possible behaviours of a space-time plate. Which solution
describes the real Universe? A closely related question is: which quantities
should be measured to verify the theory?

Note, first, that the basic object of the theory,  the  strain vector
$\overrightarrow{\xi}$ components, are not measurable by purely 4-dimensional
geometrical methods.
Moreover, all plates whose surfaces differ from each other only by an isometric
deformation are equivalent from the viewpoint of the internal local geometry (but,
generally speaking, can  describe different topology and physics). So,
measurable geometrical quantities are the same, that in GR: metric,
connection, curvature. The quantities $\xi^m$ appear as superpotentials for the metric
(gravitational field).

It would be very important to measure the elastic constants:  $E$ (the Young
modulus), $\sigma$ (Poisson coefficient), and the thickness $h_m$ of the
space-time plate. In experiments with usual 2-dimensional plates these
values can be easily found: one should study the plate
surface displacement in 3-dimensional space under given (controlled by the experimentalist)
bending forces and boundary conditions. In our case the plate under study is
space-time itself, so  such active experiments are impossible: we are able
to  control neither the bending forces, which have, generally speaking, an extra-dimensional
nature, nor the boundary conditions --- they are unknown and unique for the
Universe. However, even in our simplest models without external
multidimensional forces it is easy to note, that the elastic constant $f$
explicitly appears in the solutions under certain boundary conditions (see
(\ref{gsolfiii}),(\ref{gffii}), (\ref{gpfii})).

\begin{figure}[htb]
\centering
\input{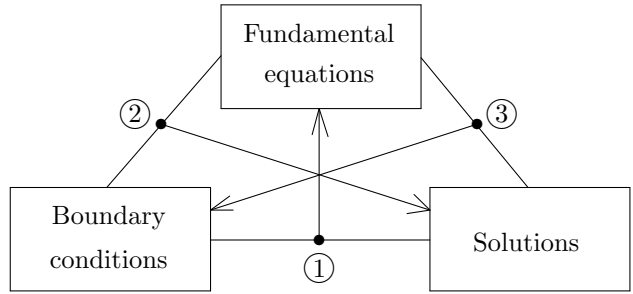}
\caption{\small Three "whales"\, of modern physics and their interrelations.}
\label{whales}
\end{figure}

But which boundary conditions should one take? Here it would be to the point to
adduce the following general scheme (fig.\ref{whales}), showing what we are
dealing with in physics.

The scheme shows three "whales"\, of any physical theory: fundamental
equations (a dynamical principle), boundary (initial) conditions and solutions.
 The first arrow shows, how a fundamental theory is born: the  dynamics and boundary
(initial) conditions observable from
 experiment (when such experiments are possible!)
 allow to treat the evolution of the system as a solution to some set of
 fundamental equations.  This restoration of the theory is very often not
 straightforward and unique. The second arrow shows the classical problem of
 fundamental theory: having the fundamental equations and boundary conditions we
 try to obtain solutions and then to compare them with experimental (observable)
 data. The existence and uniqueness theorems in the corresponding  mathematical
 theory testify that this problem is not trivial too.

 And, lastly, the third arrow shows remaining non-traditional way of
 investigation, when the boundary conditions are unknown: using the fundamental equations
 and comparing their solution with the observable system evolution law, one
 can obtain some information (perhaps incomplete and non-unique) about the boundary
 conditions of the system. This last case does take  place, when we study the
global properties
 of Universe. So, in cosmological problems we should use the observational data
 not only to test our dynamical (say, Einstein) equations, but also for
an  indirect investigation of the bounds of the Universe.
In particular, experimental measurements of one or several 4-dimensional
geometrical objects (say, invariants of the  curvature tensor)\footnote{In our
model there is one independent quantity, the curvature scalar: ${}^2R(x,t)$.}
and their comparison with the theoretical ones can be used for identification of
the boundary conditions of the Universe.

To illustrate this idea, let us compare the behaviour of the curvature scalar
${}^2R_{obs}$, (taken on the 1-dimensional past light cone $x\pm t=0$) for some
of the above examples with different boundary conditions.

1) Example
(\ref{eiiii}) with the condition $(\infty,\infty,\infty,\infty)$  gives
\begin{equation}\label{riiii}
{}^2R_{obs}={}^2R|_{t=x}=\frac{1}{4}(x^2\cos2x-x\sin2x+\frac{1}{4}\sin^22x).
\end{equation}
It is shown in Fig.\ref{friiii}.

2) Example (\ref{esiii}) with the condition $(S,\infty,\infty,\infty)$ gives
\begin{equation}\label{rsiii}
{}^2R_{obs}=2x\sin x+3\cos 2x+\cos^4x.
\end{equation}
It is shown in Fig.\ref{frsiii}.

3)
Example (\ref{epsii}) with the condition $(P,S,\infty,\infty)$
 gives
\begin{equation}\label{rpsii}
{}^2R_{obs}=\cos^4x-\cos 2x.
\end{equation}
It is shown in Fig.\ref{frpsii}.

4) Example (\ref{essii}) with the condition $(S,S,\infty,\infty)$ gives
\begin{equation}\label{rssii}
{}^2R_{obs}=4\cos^4x-3\cos2x.
\end{equation}
It is shown in Fig.\ref{frssii}

5) Example (\ref{effii}) with the condition  $(F,F,\infty,\infty)$ gives
\begin{equation}\label{rffii}
{}^2R_{obs}=6400x^8.
\end{equation}
It is plotted (in a smaller scale) in Fig.\ref{frffii}.

6)  Example (\ref{essss}) with the condition $(S,S,S,S)$ gives
\begin{equation}\label{rssss}
{}^2R_{obs}=\cos2x.
\end{equation}
It is plotted in Fig.\ref{frssss}.

From (\ref{riiii})-(\ref{rssss}) and Figs.\ref{friiii}-\ref{frssss} it is
easy to see that different boundary conditions, generally speaking, lead to
different types of behaviour of the plate and partially enables one  to distinguish the
conditions from each other by experiment at least in principle.

As follows from Sections 7,8, the presence of two opposite bounds leads to a discrete
spectrum of standing waves. Its form can be expressed by the following formula:
\begin{equation}\label{spectr}
\omega_{n}=\frac{\pi n}{\tau},\ n\in \mathbb{N},
\end{equation}
$\tau$
being  the unstrained plate length  in the time (or space) direction. So, if the measured
${}^4R_{exp}$ is periodical, then Fourier analysis will give us the set
$\{\omega_n\}$, where the information about the size of the Universe is hidden. In the
elastic picture it is reasonable that the periodical dependence of geometric
objects correlates with the matter distribution periodicity. Then the observable
cellular structure of the Universe gets a natural explanation in our approach: the spatial cells
and matter walls are a 3-dimensional projection of a standing wave on the
space-time plate, which is formed due to the presence of bounds (or, more
generally, due to a finite volume of the Universe).

In our model, periodicities in space and time are correlated, so we can get an estimate of the
time length of the Universe. The mean size $\lambda$ of a space cell
is about $30$MPc. Using (\ref{spectr}), we find $\tau\sim (\lambda/c)n$. Since
$n\ge1$, then $\tau\ge\lambda/c\sim10^7$ y, which does not contradict to the data
on present age of the Universe, $\sim10^{10}$ y.

One can conclude that purely geometric measurements
do not give full information about the elastic properties of
the plate Universe: it is necessary to supplement
geometrical measurements by  physical experiments.
For example, it is intuitively clear that the usual Young modulus
and Poisson coefficient of 4-dimensional matter\footnote{
The energy density of 4-dimensional matter is a pressure
of space-time substance in time-like direction.}
are the corresponding "reduced"\, multidimensional
constants. If the thickness $h_m$ of the plate may be related to
quantum phenomena, then it can be determined from atomic, nuclear
or elementary particle physics.

\section{Conclusion}

The above analysis shows that,  first of all, in spite of the analogy with
the theory of deformation of common plates, the solutions of equilibrium
equations of the 4D plates have some unusual  properties. Namely, for the
above models the linear "swinging"\, of a displacement vector along
time and correspondingly quadratic "swinging"\, of the  curvature scalar are typical
(see Figs.\ref{fiiii}-\ref{frsiii}) . The origin of this swinging are high
derivatives of the biwave equation. Note, that this swinging naturally leads to a
more general strong bending theory.

The presence of bounds and their type are, in principle, experimentally testable by
analysis of observable geometrical object (or, indirectly, matter
distribution). The presence of opposite bounds gives some discrete set of
frequences, which can be used for calculation of the space and time size of
Universe.

One should stress that the above  solutions are nothing more than "toy"\, models, possessing
some typical peculiarities of the suggested approach. More realistic
models can be obtained in a more general context than the linear
theory, which has been developed in \cite{kok3}.

\vspace{2cm}
{\large\bf Aknowlegments}

I am most grateful to V.A. Korotky for useful discussions, and E.P. Shtern for
technical support.

\newpage

\onecolumn
{\parindent1cm\unitlength=\textwidth
\refstepcounter{figure}\label{fiiii}
\begin{picture}(0.45,0.4)
\put(-0.04,0.4){\includegraphics[width=0.4\textwidth,angle=-90]{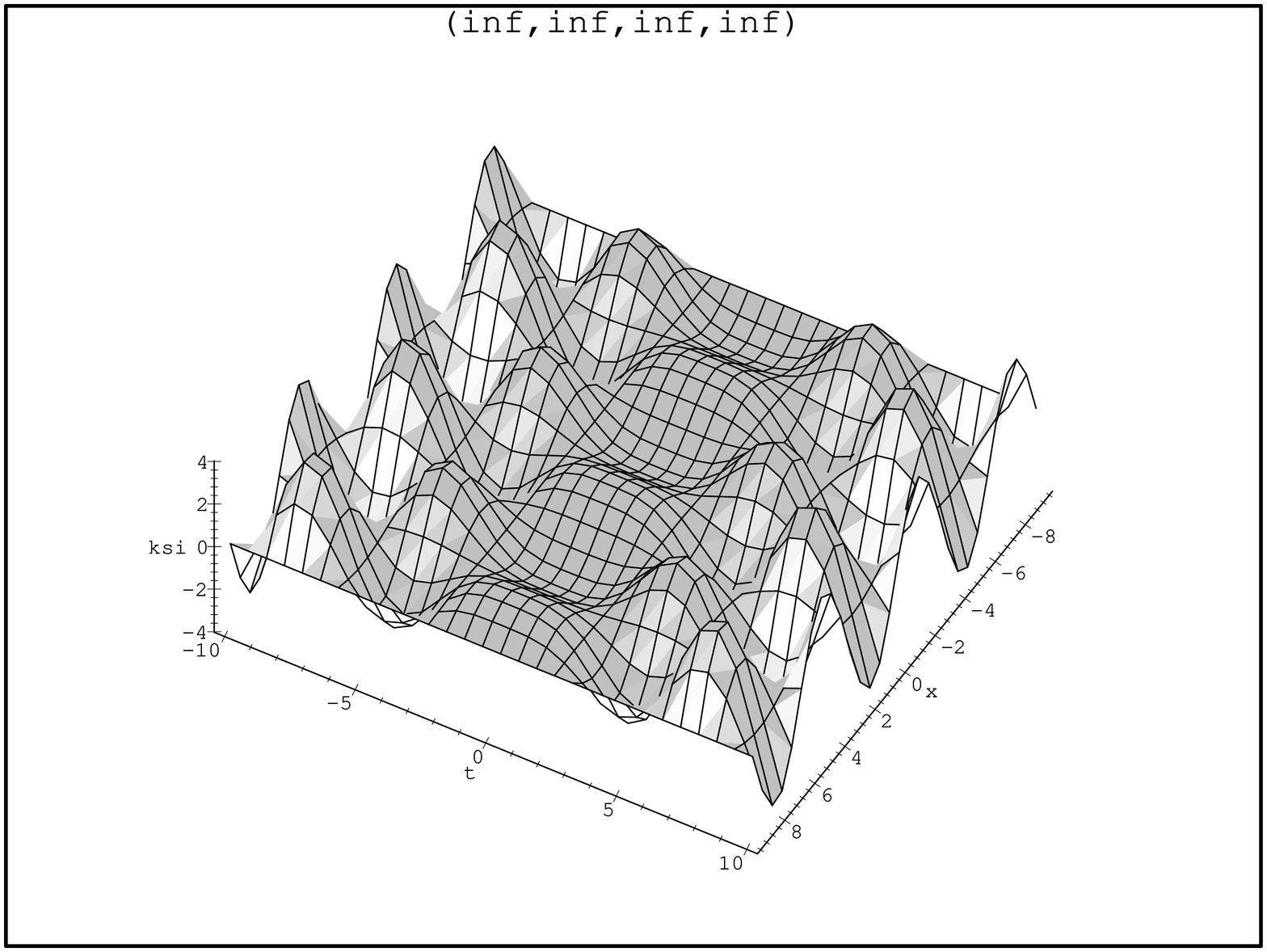}}
\put(0.0,0){\makebox(0.4,0)[cc]{{\thefigure:~$(\infty,\infty,\infty,\infty)$-plate.}}}
\end{picture}
\hfill \refstepcounter{figure}\label{fpiii}
\begin{picture}(0.45,0.4)
\put(-0.04,0.4){\includegraphics[width=0.4\textwidth,angle=-90]{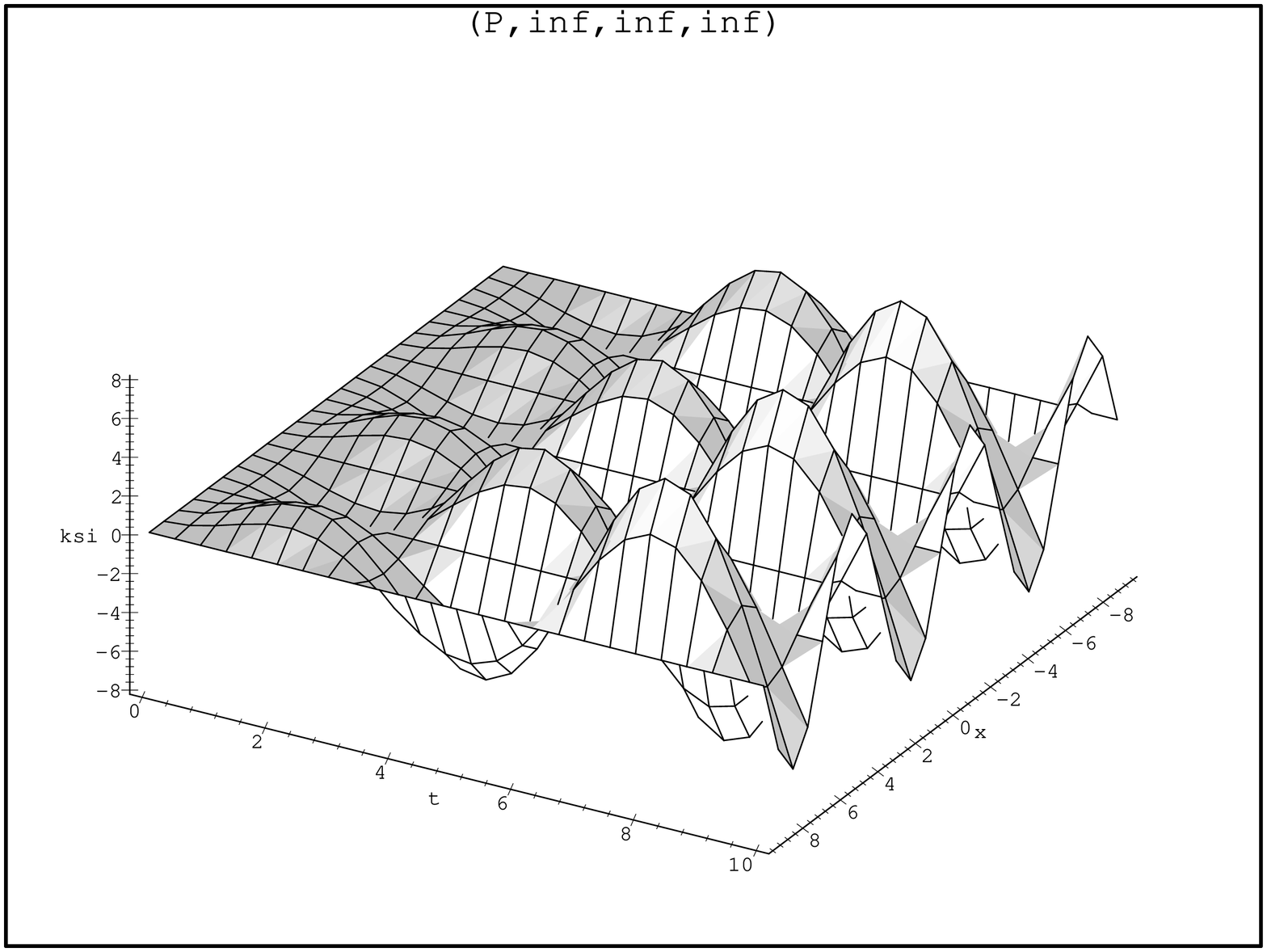}}
\put(0.0,0){\makebox(0.4,0)[cc]{{\thefigure:~$(P,\infty,\infty,\infty)$-plate.}}}
\end{picture}

\vspace{0.5cm}
\parindent1cm
\refstepcounter{figure}\label{fsiii}
\begin{picture}(0.45,0.4)
\put(-0.04,0.4){\includegraphics[width=0.4\textwidth,angle=-90]{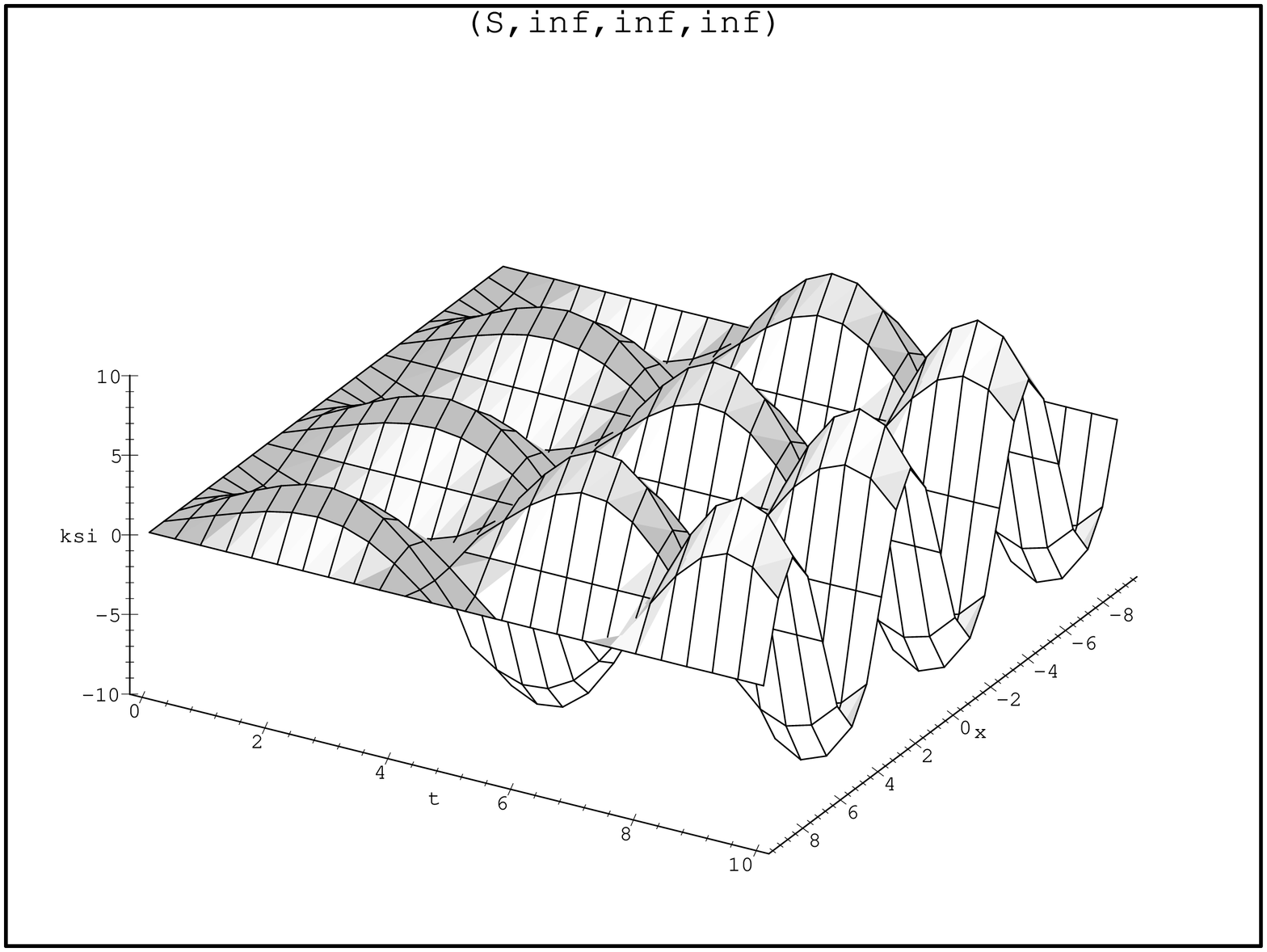}}
\put(0.0,0){\makebox(0.4,0)[cc]{{\thefigure:~$(S,\infty,\infty,\infty)$-plate.}}}
\end{picture}
\hfill \refstepcounter{figure}\label{ffiii}
\begin{picture}(0.45,0.4)
\put(-0.04,0.4){\includegraphics[width=0.4\textwidth,angle=-90]{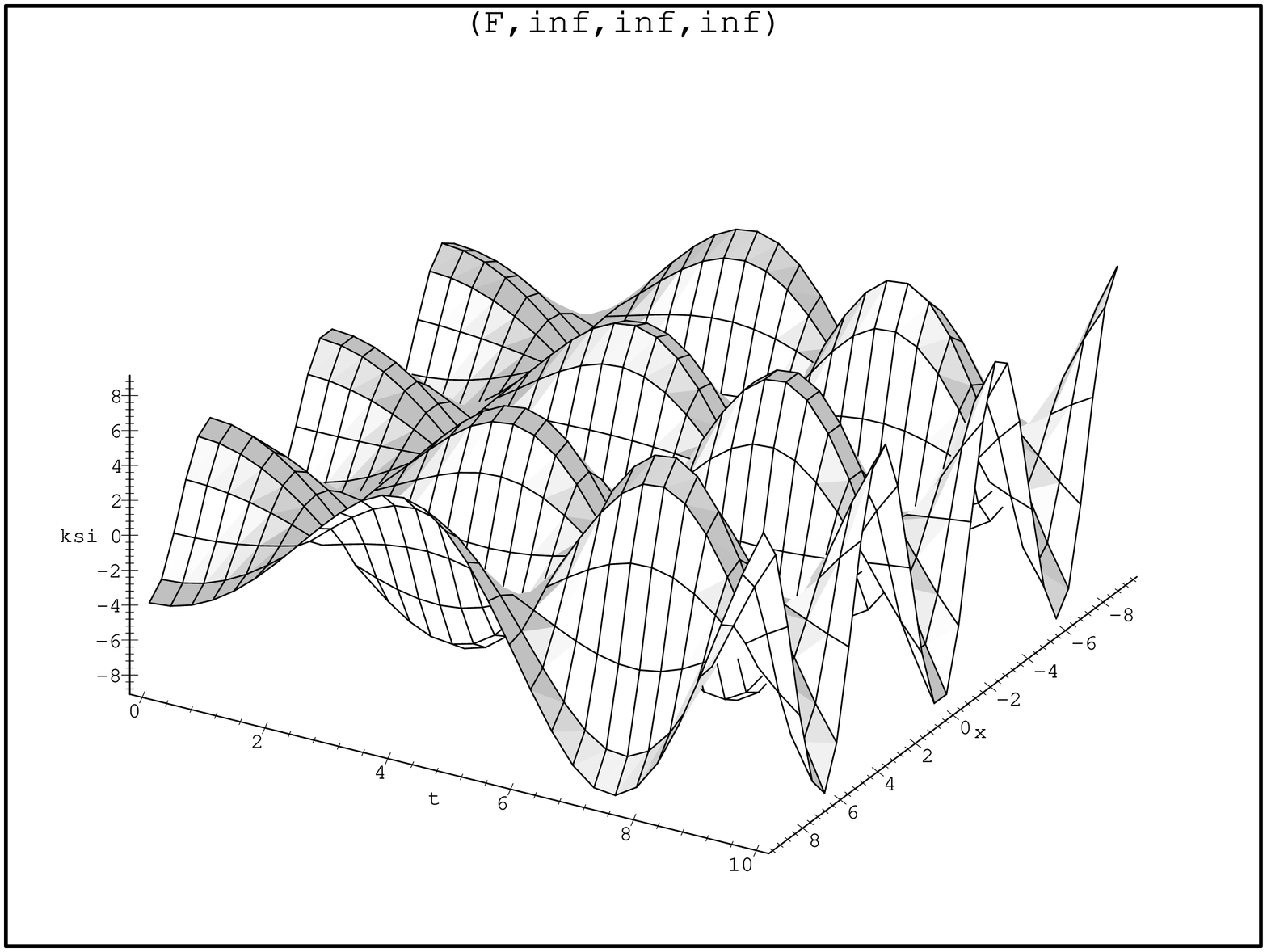}}
\put(0.0,0){\makebox(0.4,0)[cc]{{\thefigure:~$(F,\infty,\infty,\infty)$-plate.}}}
\end{picture}

\vspace{0.5cm}
\parindent1cm\refstepcounter{figure}\label{fssii}
\begin{picture}(0.45,0.4)
\put(-0.04,0.4){\includegraphics[width=0.4\textwidth,angle=-90]{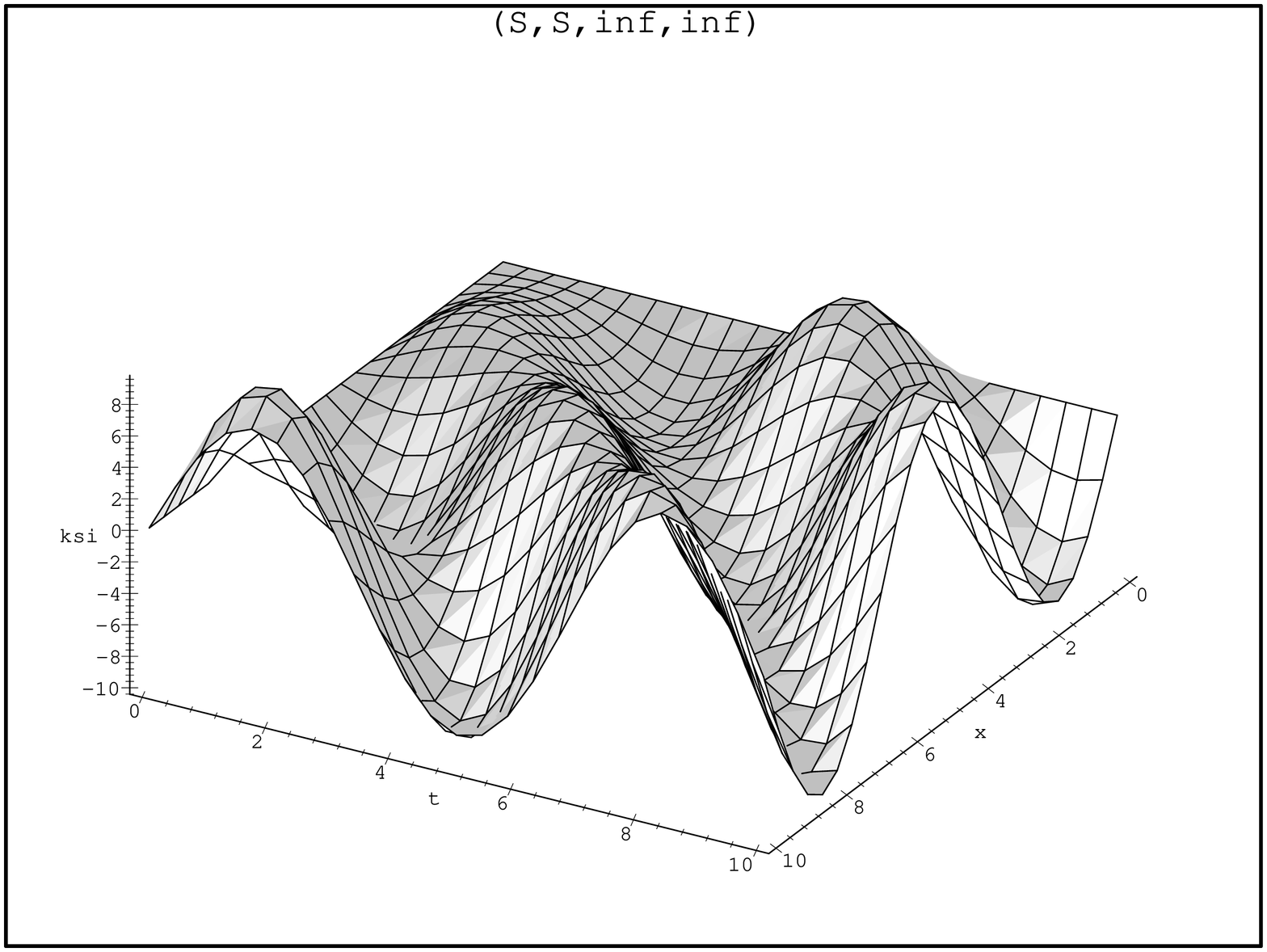}}
\put(0.0,0){\makebox(0.4,0)[cc]{{\thefigure:~$(S,S,\infty,\infty)$-plate.}}}
\end{picture}
\refstepcounter{figure}\label{fffii}
\begin{picture}(0.45,0.4)
\put(-0.04,0.4){\includegraphics[width=0.4\textwidth,angle=-90]{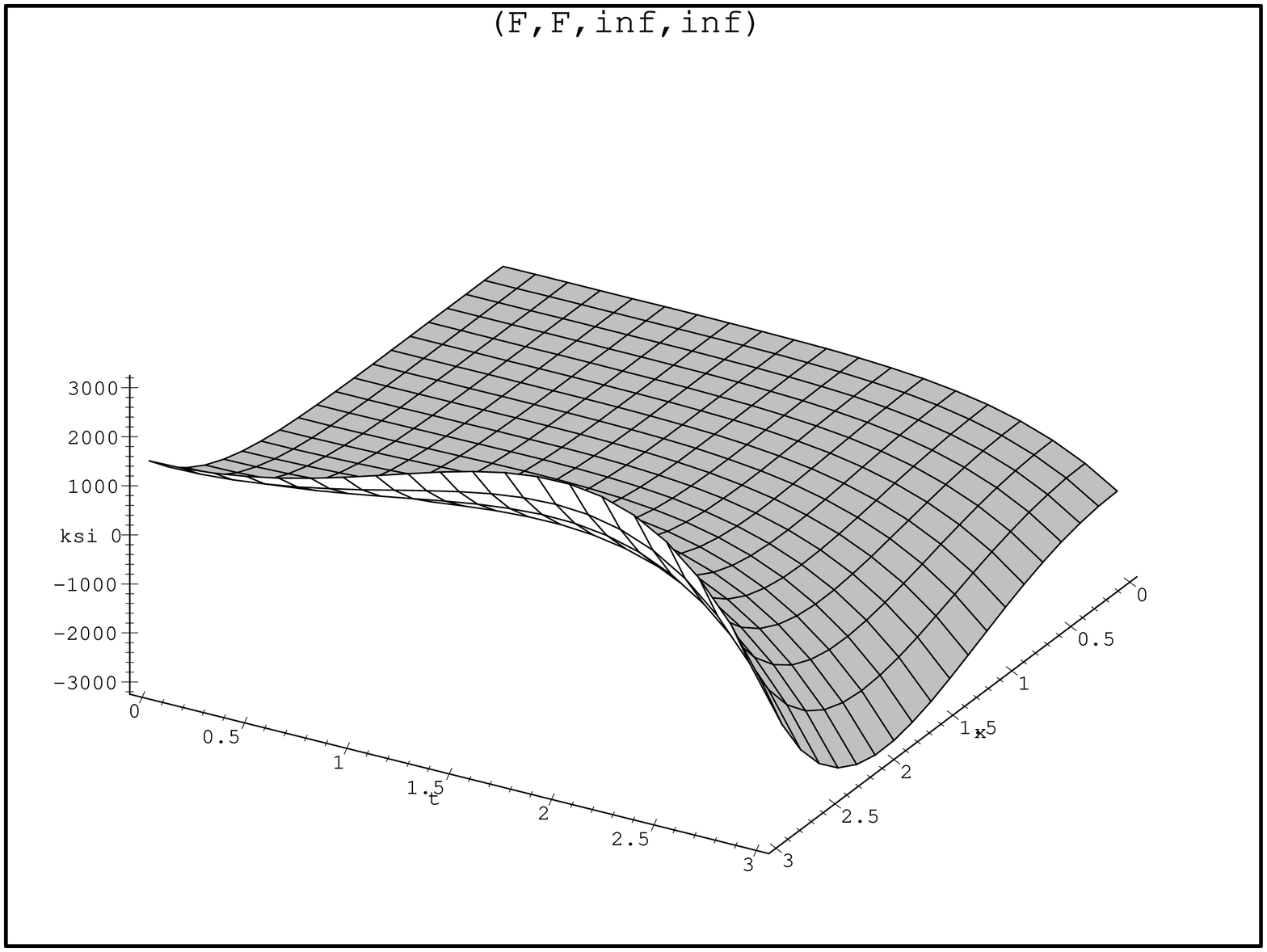}}
\put(0.0,0){\makebox(0.4,0)[cc]{{\thefigure:~$(F,F,\infty,\infty)$-plate.}}}
\end{picture}
}

\newpage

\onecolumn {\parindent1cm\unitlength=\textwidth
\refstepcounter{figure}\label{fpsii}
\begin{picture}(0.45,0.4)
\put(-0.04,0.4){\includegraphics[width=0.4\textwidth,angle=-90]{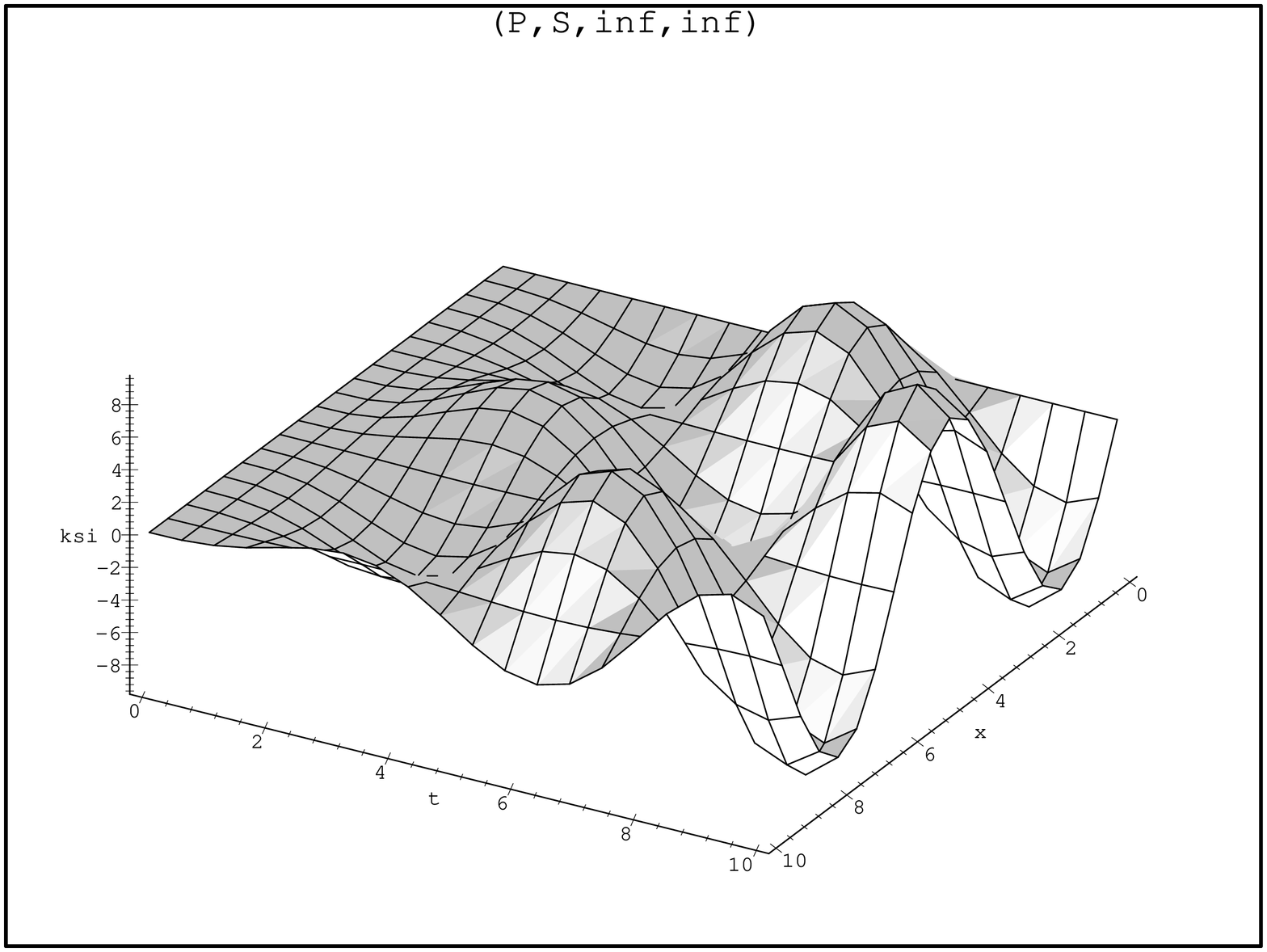}}
\put(0.0,0){\makebox(0.4,0)[cc]{{\thefigure:~$(P,S,\infty,\infty)$-plate.}}}
\end{picture}
\hfill \refstepcounter{figure}\label{fpfii}
\begin{picture}(0.45,0.4)
\put(-0.04,0.4){\includegraphics[width=0.4\textwidth,angle=-90]{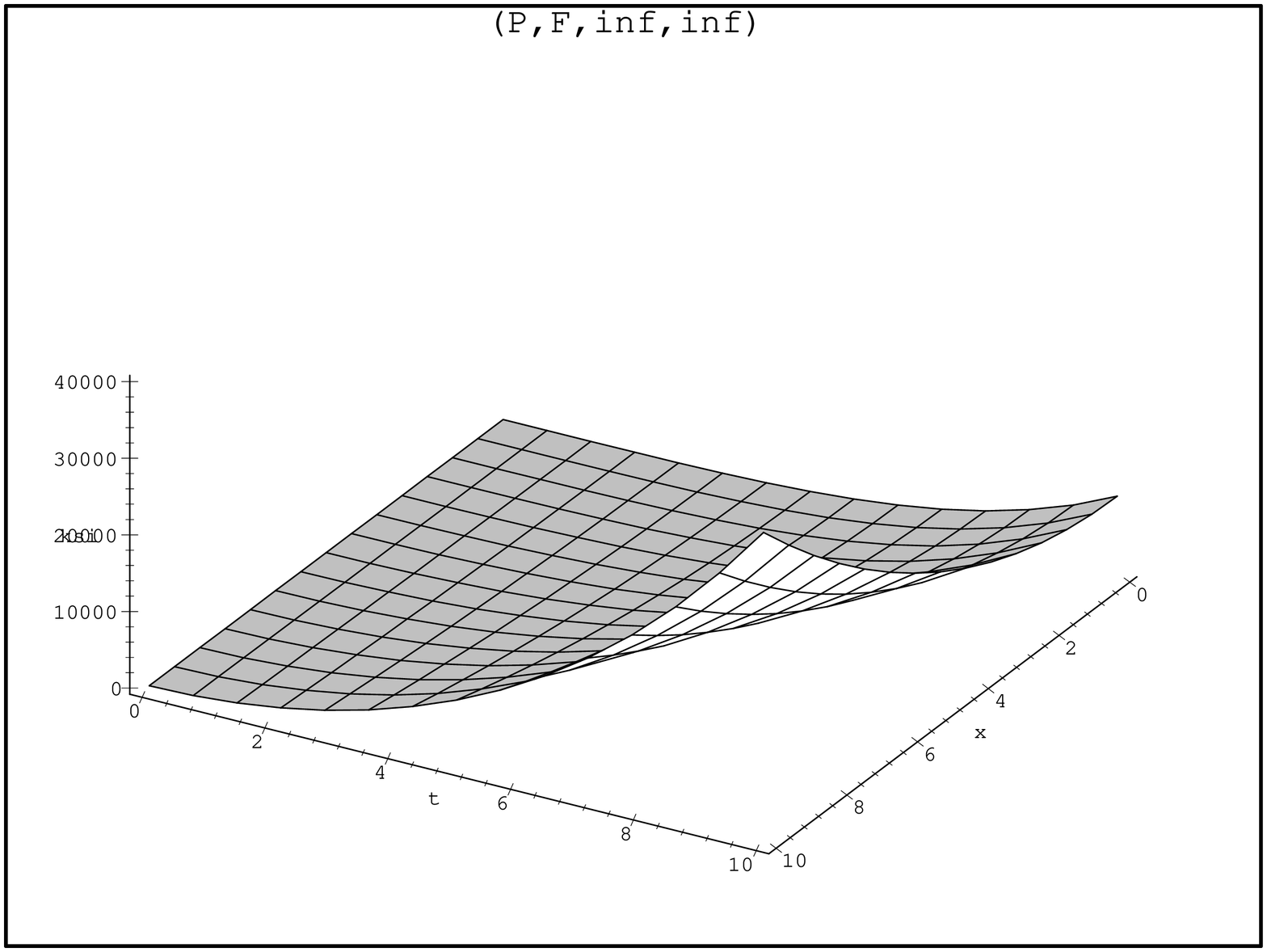}}
\put(0.0,0){\makebox(0.4,0)[cc]{{\thefigure:~$(P,F,\infty,\infty)$-plate.}}}
\end{picture}

\vspace{1cm}
\parindent1cm\refstepcounter{figure}\label{ffsii}
\begin{picture}(0.45,0.4)
\put(-0.04,0.4){\includegraphics[width=0.4\textwidth,angle=-90]{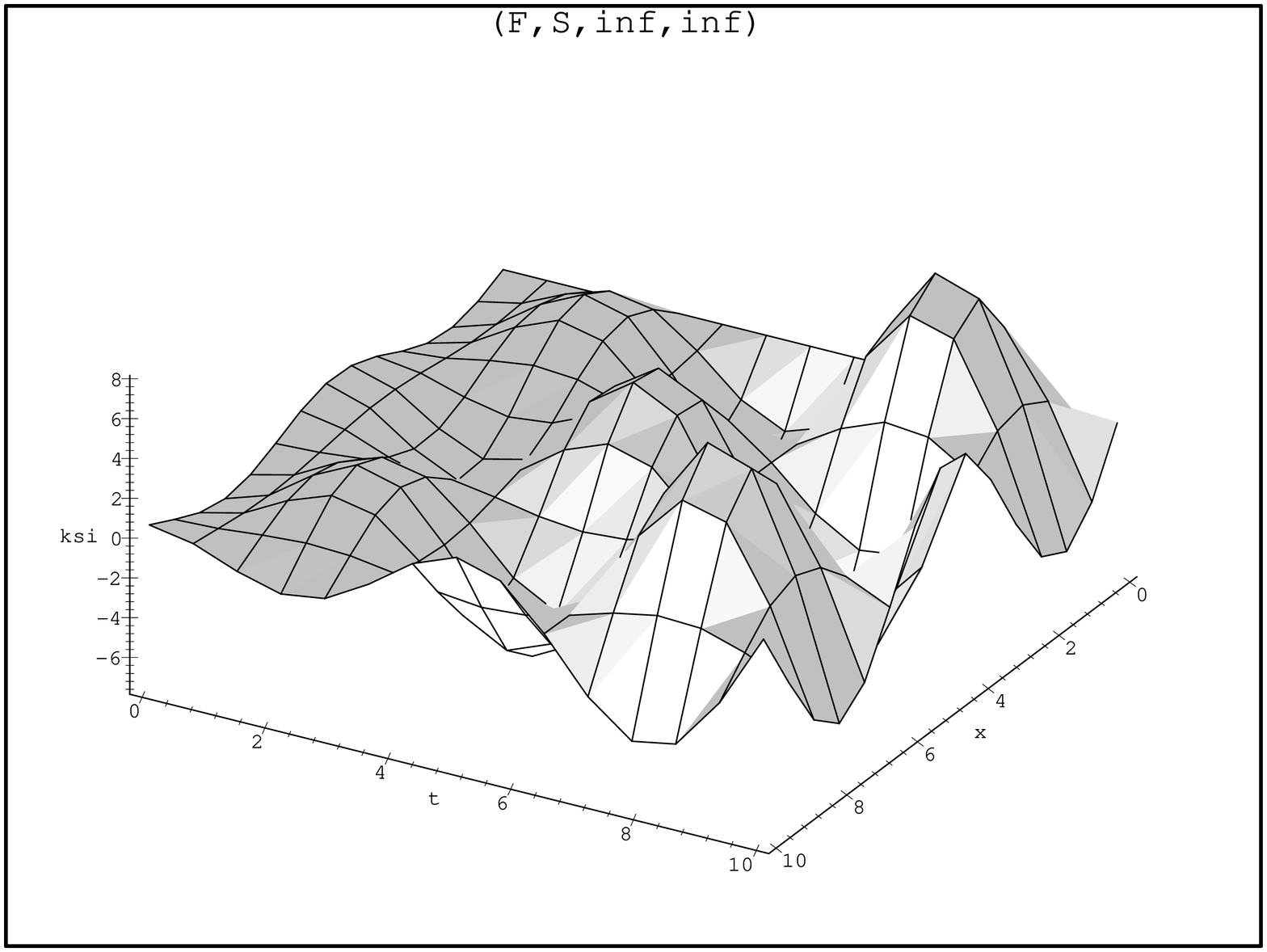}}
\put(0.0,0){\makebox(0.4,0)[cc]{{\thefigure:~$(F,S,\infty,\infty)$-plate.}}}
\end{picture}
\hfill \refstepcounter{figure}\label{fsisi}
\begin{picture}(0.45,0.4)
\put(-0.04,0.4){\includegraphics[width=0.4\textwidth,angle=-90]{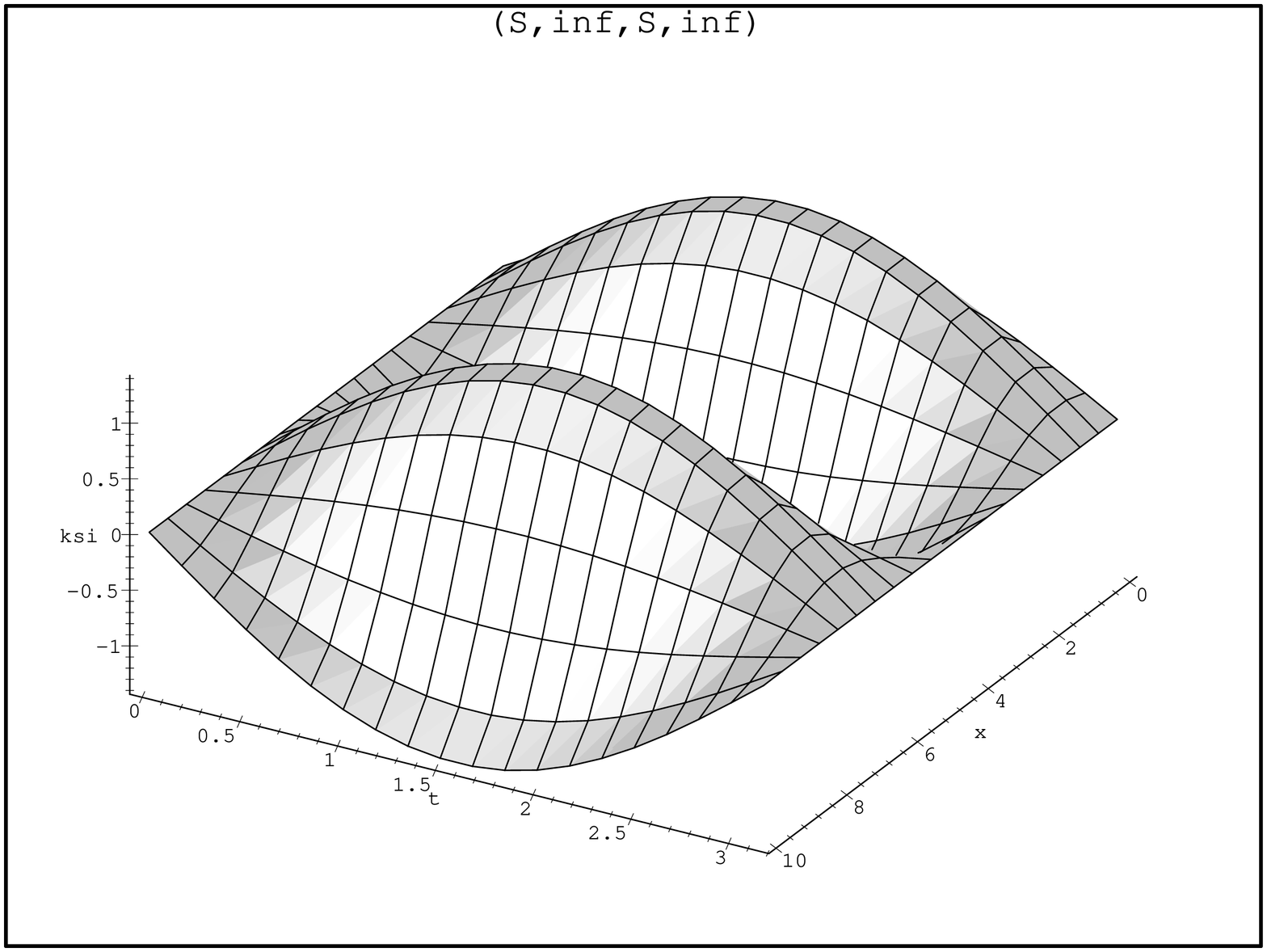}}
\put(0.0,0){\makebox(0.4,0)[cc]{{\thefigure:~$(S,\infty,S,\infty)$-plate.}}}
\end{picture}

\vspace{1cm}
\parindent1cm\refstepcounter{figure}\label{fsssi}
\begin{picture}(0.45,0.4)
\put(-0.04,0.4){\includegraphics[width=0.4\textwidth,angle=-90]{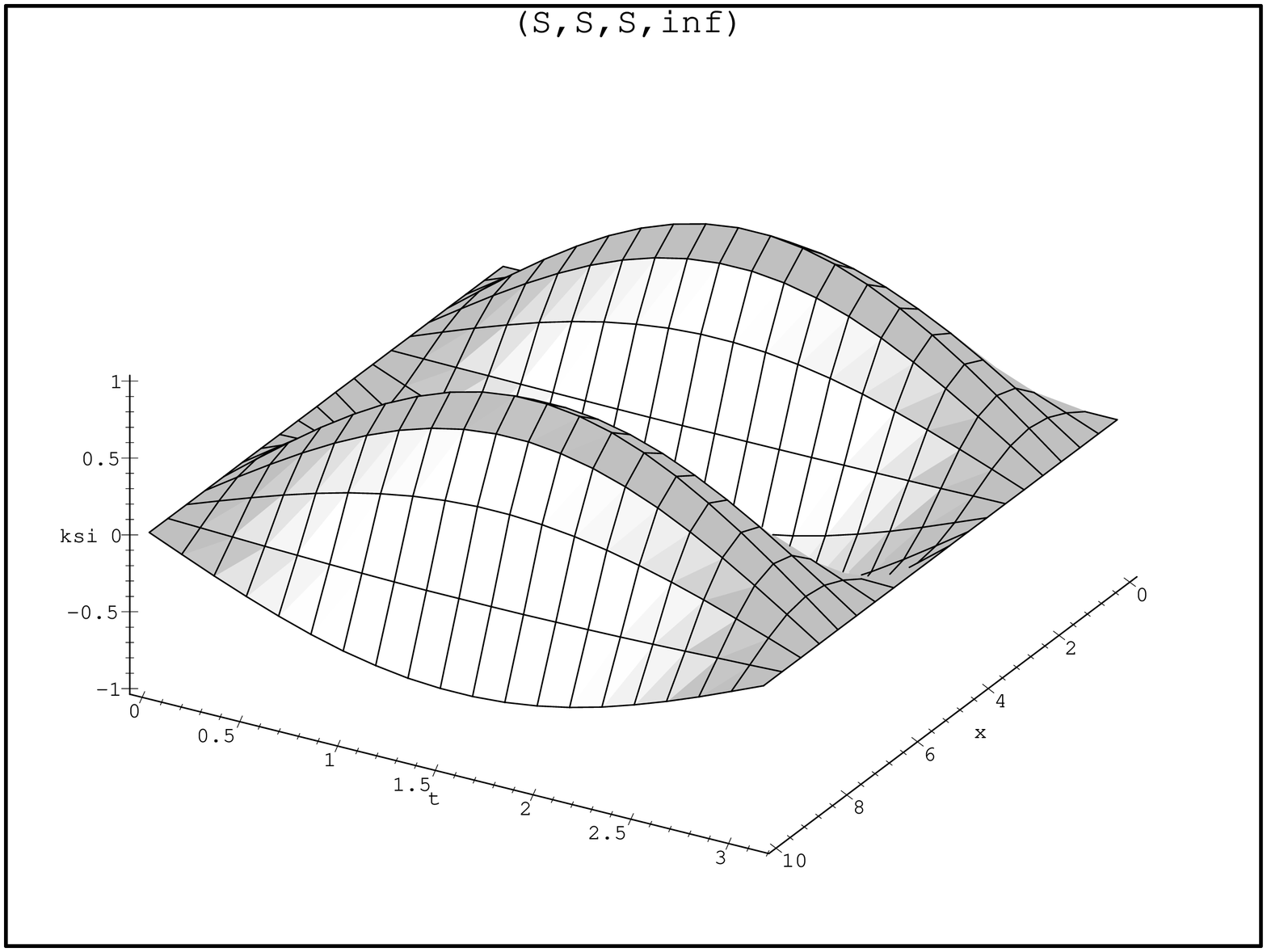}}
\put(0.0,0){\makebox(0.4,0)[cc]{{\thefigure:~$(S,S,S,\infty)$-plate.}}}
\end{picture}
\refstepcounter{figure}\label{fssss}
\begin{picture}(0.45,0.4)
\put(-0.04,0.4){\includegraphics[width=0.4\textwidth,angle=-90]{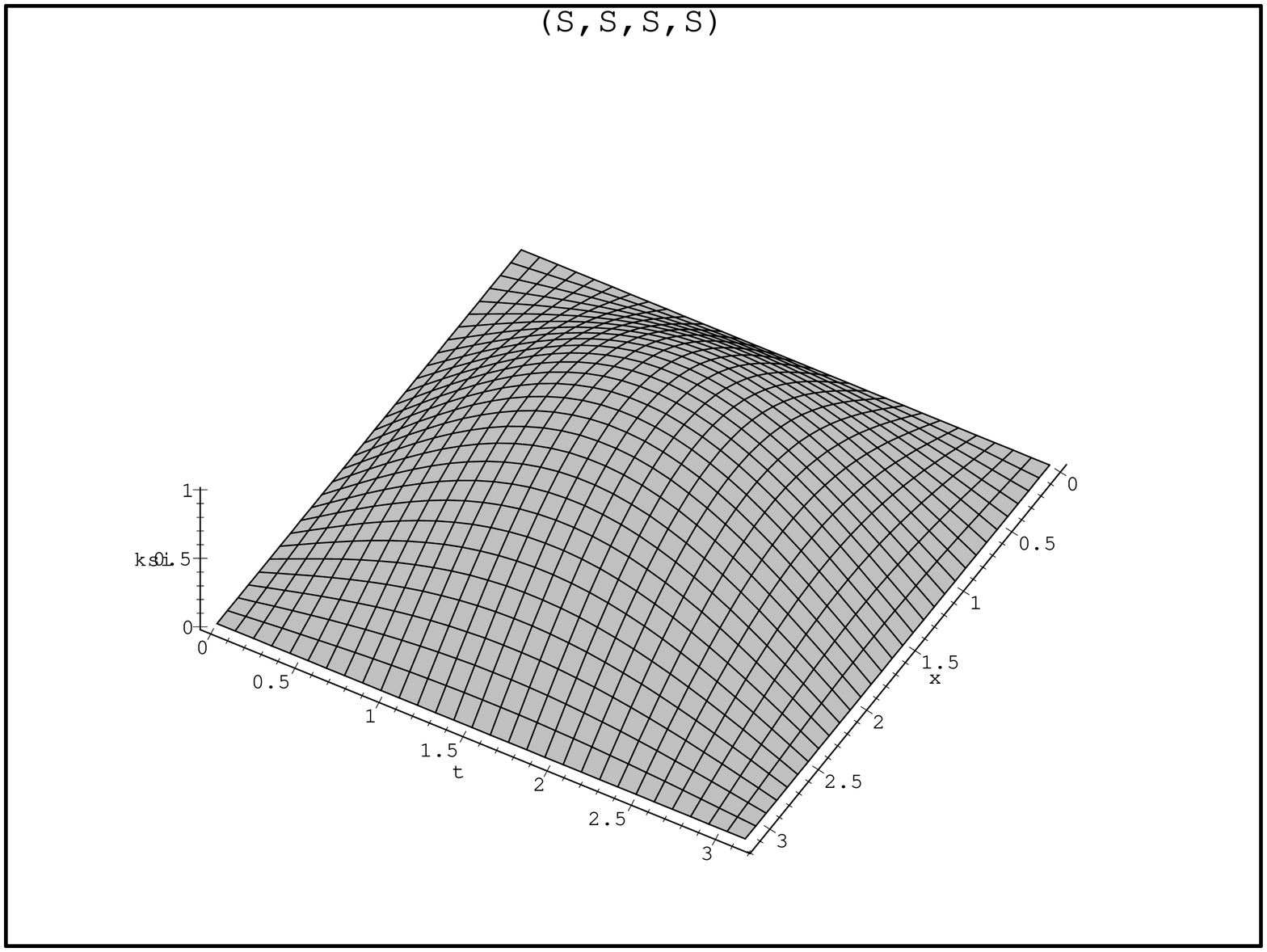}}
\put(0.0,0){\makebox(0.4,0)[cc]{{\thefigure:~$(S,S,S,S)$-plate.}}}
\end{picture}
}
\newpage

\onecolumn {\parindent1cm\unitlength=\textwidth
\refstepcounter{figure}\label{friiii}
\begin{picture}(0.45,0.4)
\put(-0.04,0.4){\includegraphics[width=0.4\textwidth,angle=-90]{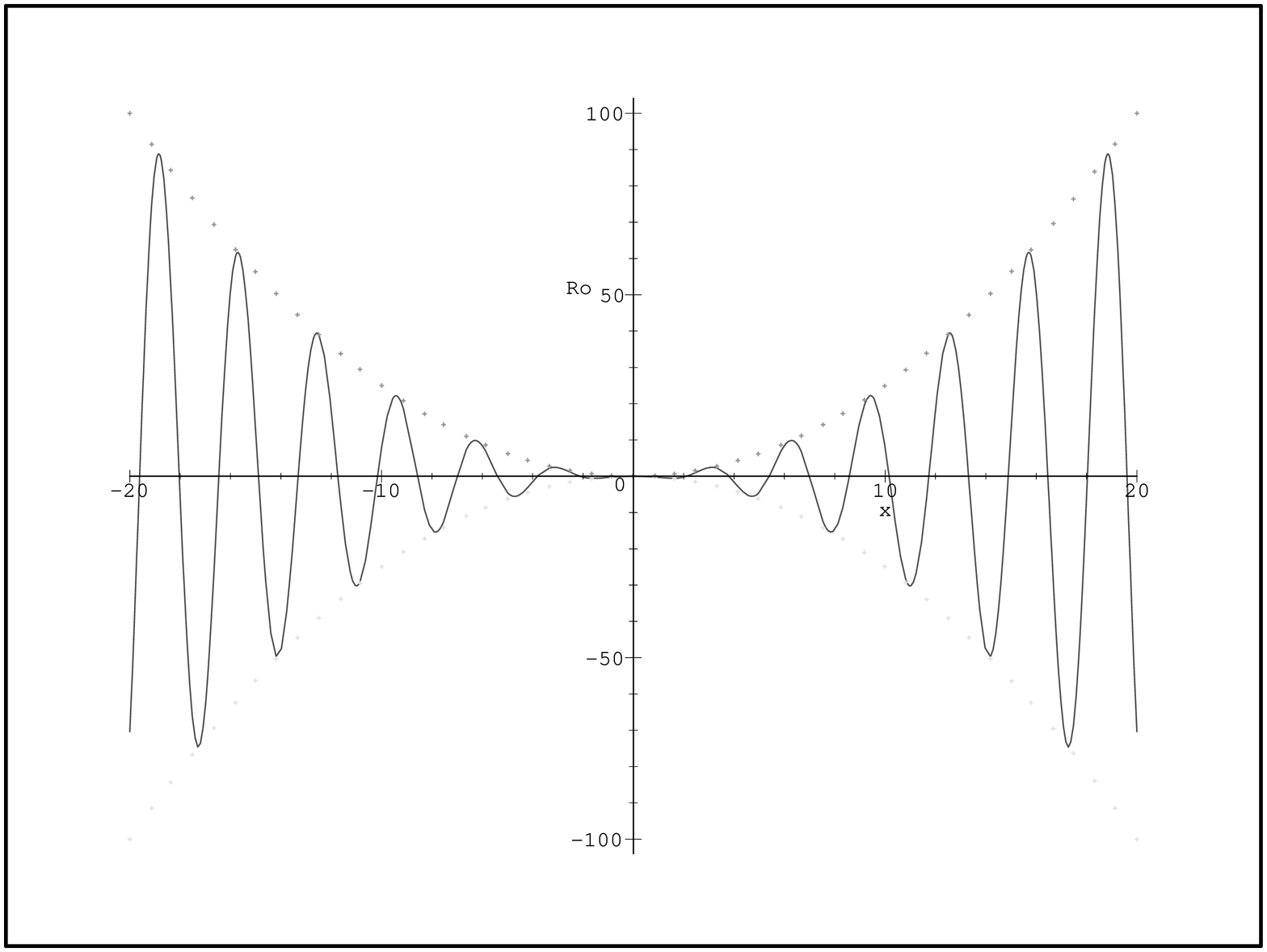}}
\put(0.0,0){\makebox(0.4,0)[cc]{{\thefigure: ${}^2R_{obs}$\ for\
$(\infty,\infty,\infty,\infty)$-plate.}}}
\end{picture}
\hfill \refstepcounter{figure}\label{frsiii}
\begin{picture}(0.45,0.4)
\put(-0.04,0.4){\includegraphics[width=0.4\textwidth,angle=-90]{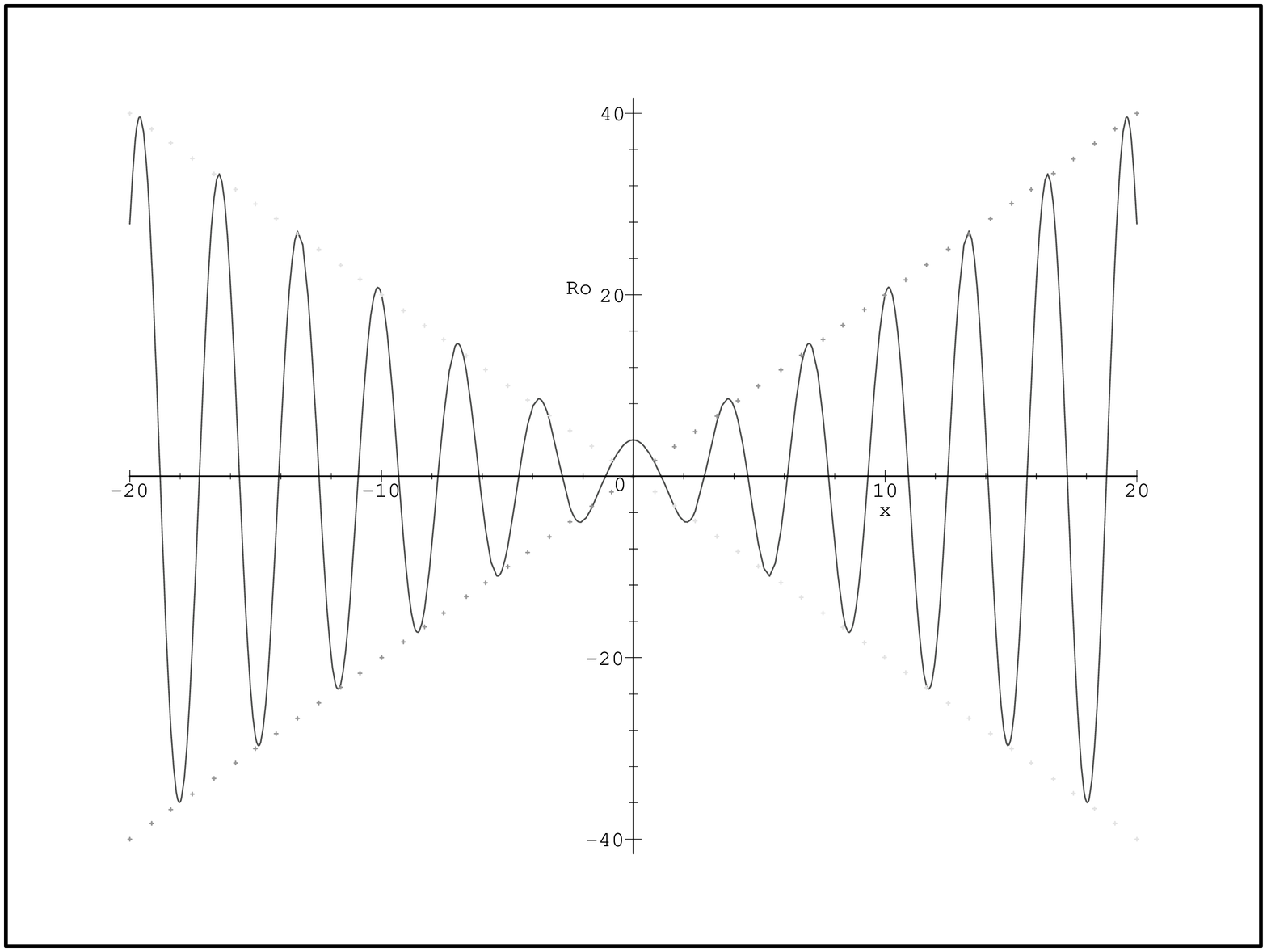}}
\put(0.0,0){\makebox(0.4,0)[cc]{{\thefigure: ${}^2R_{obs}$\ for\
$(S,\infty,\infty,\infty)$-plate.}}}
\end{picture}

\vspace{1cm}
\parindent1cm\refstepcounter{figure}\label{frssii}
\begin{picture}(0.45,0.4)
\put(-0.04,0.4){\includegraphics[width=0.4\textwidth,angle=-90]{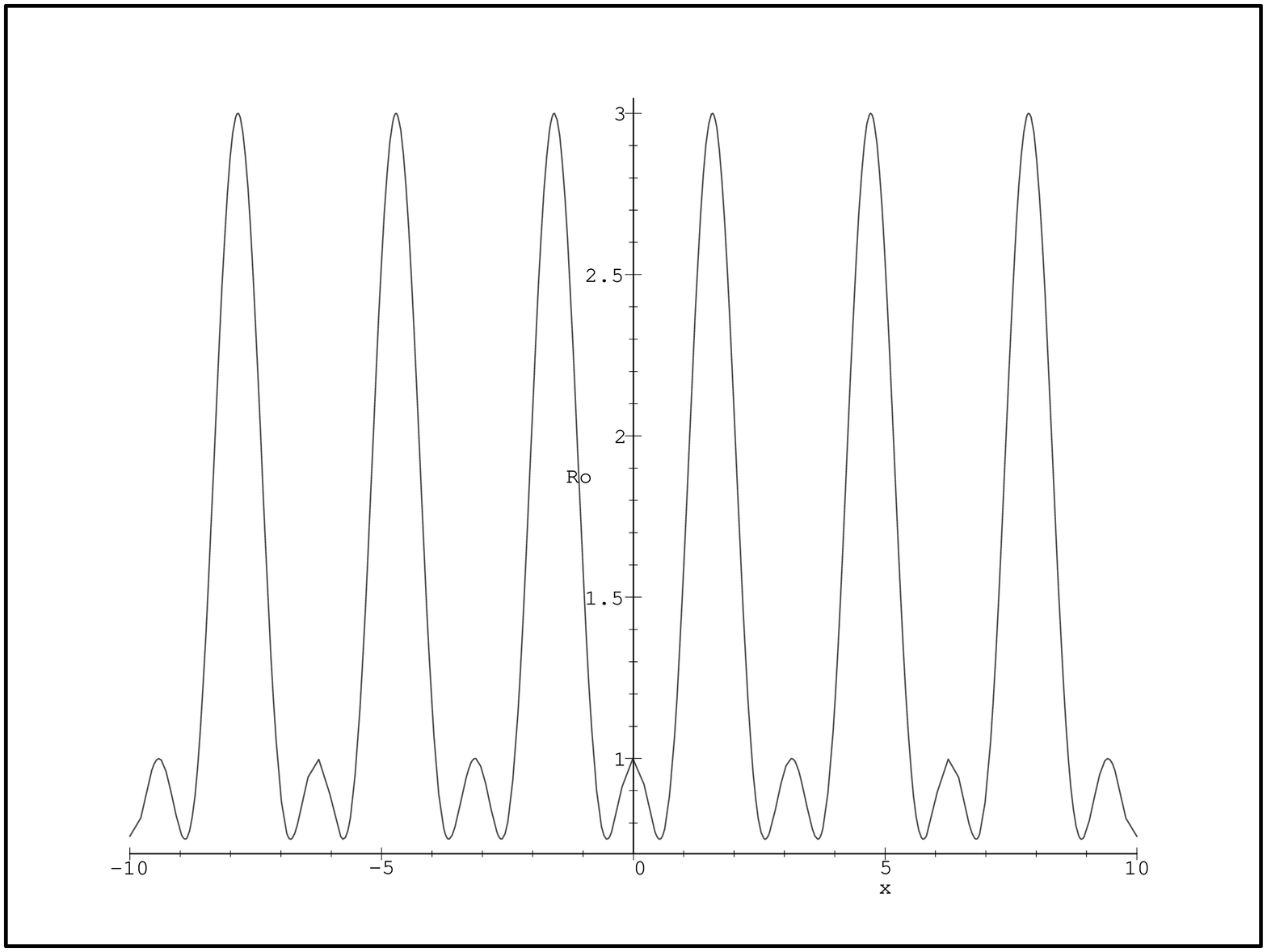}}
\put(0.0,0){\makebox(0.4,0)[cc]{{\thefigure: ${}^2R_{obs}$\ for\
$(S,S,\infty,\infty)$-plate.}}}
\end{picture}
\hfill \refstepcounter{figure}\label{frffii}
\begin{picture}(0.45,0.4)
\put(-0.04,0.4){\includegraphics[width=0.4\textwidth,angle=-90]{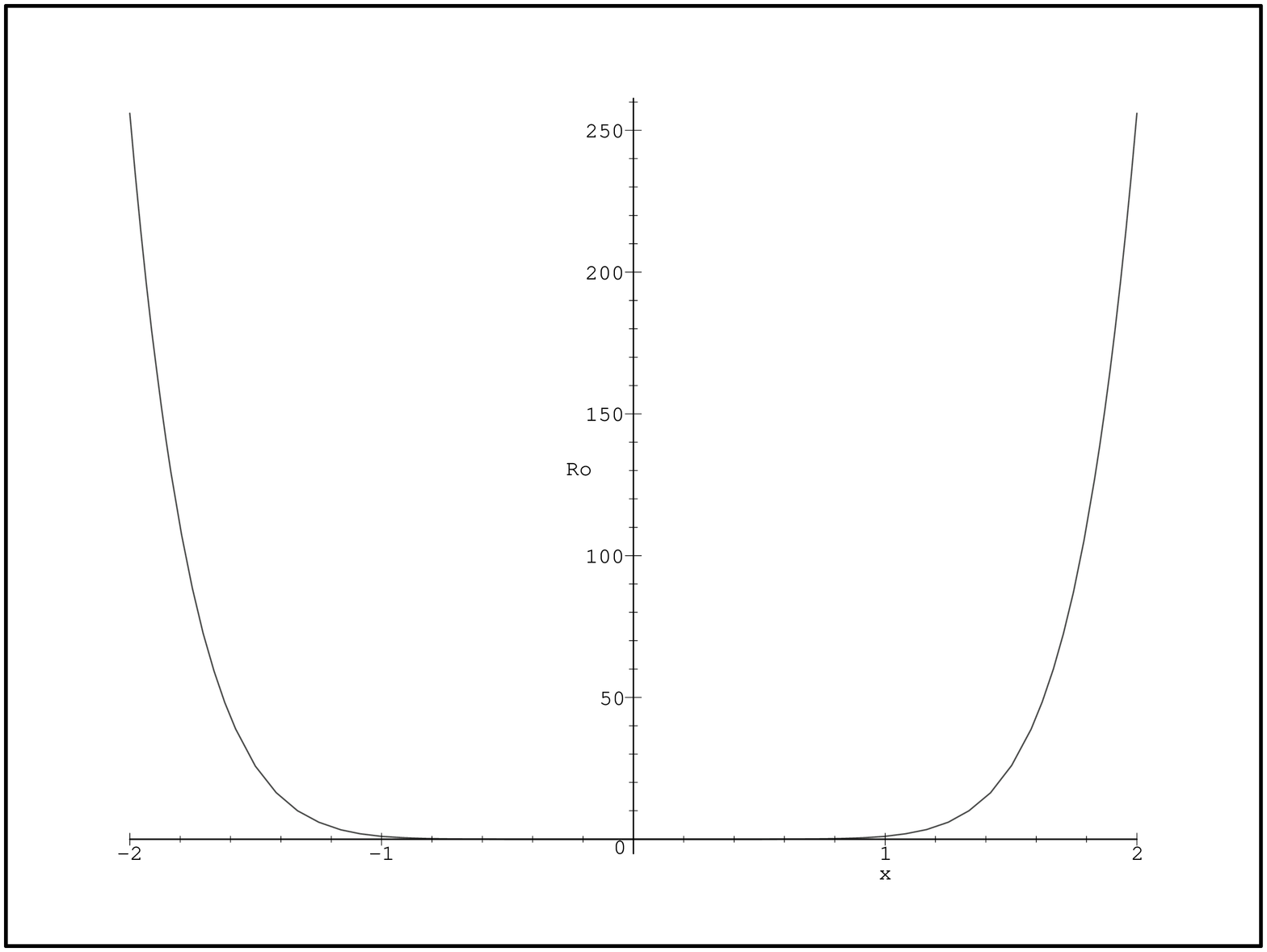}}
\put(0.0,0){\makebox(0.4,0)[cc]{{\thefigure: ${}^2R_{obs}$ for
$(F,F,\infty,\infty)$-plate.}}}
\end{picture}

\vspace{1cm}
\parindent1cm\refstepcounter{figure}\label{frpsii}
\begin{picture}(0.45,0.4)
\put(-0.04,0.4){\includegraphics[width=0.4\textwidth,angle=-90]{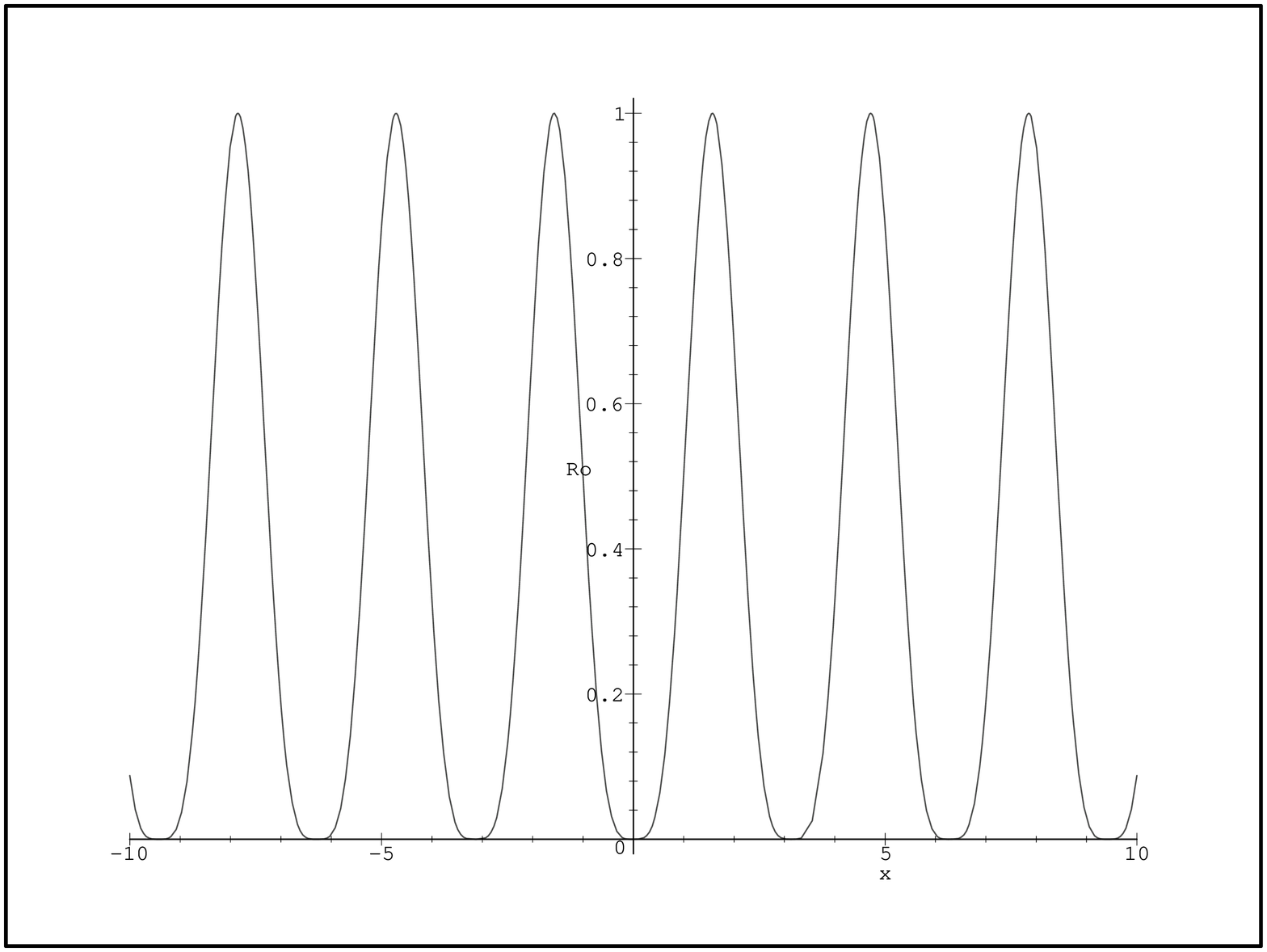}}
\put(0.0,0){\makebox(0.4,0)[cc]{{\thefigure: ${}^2R_{obs}$\ for\
$(P,S,\infty,\infty)$-plate.}}}
\end{picture}
\refstepcounter{figure}\label{frssss}
\begin{picture}(0.45,0.4)
\put(-0.04,0.4){\includegraphics[width=0.4\textwidth,angle=-90]{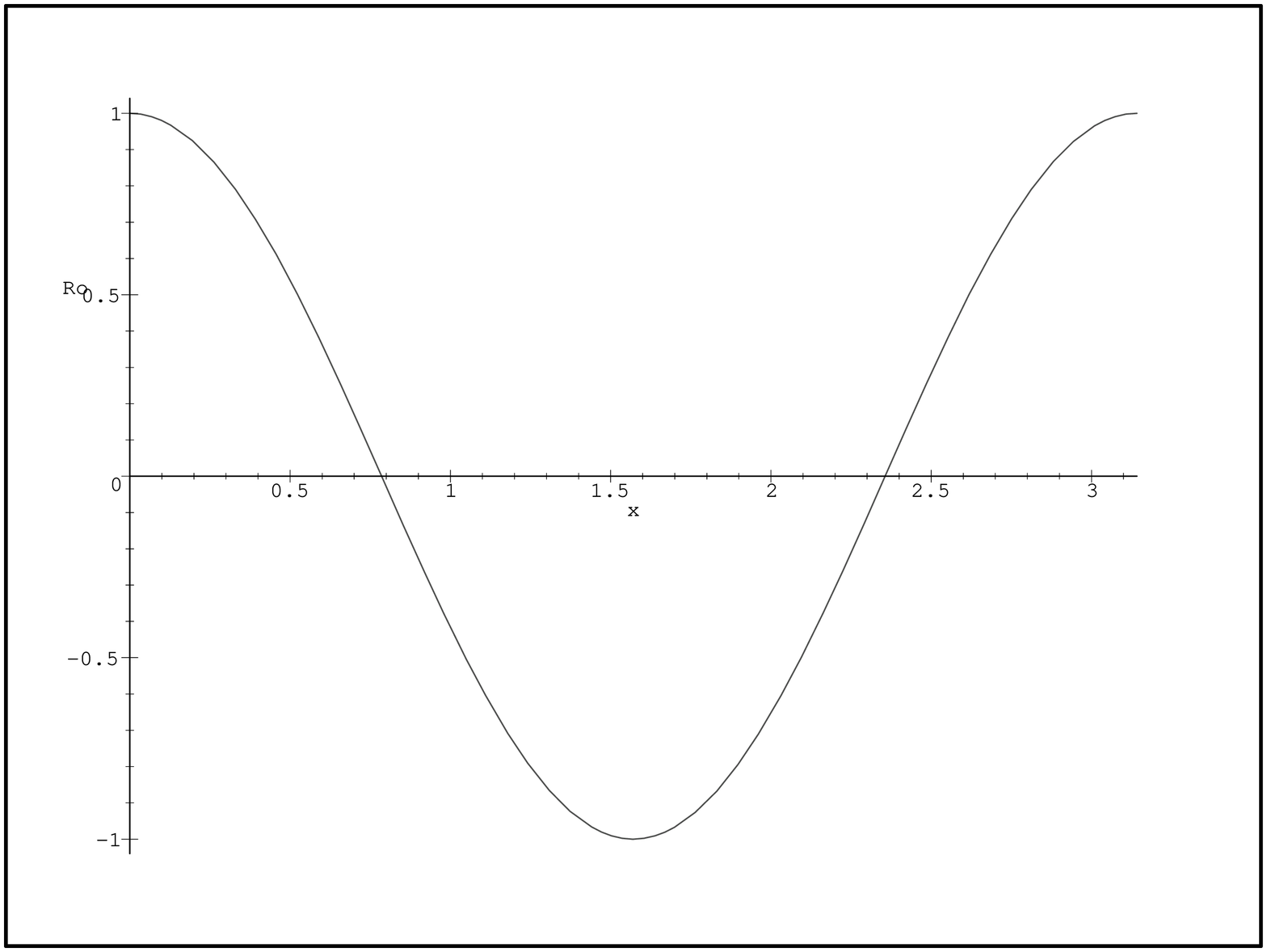}}
\put(0.0,0){\makebox(0.4,0)[cc]{{\thefigure: ${}^2R_{obs}$\ for\
$(S,S,S,S)$-plate.}}}
\end{picture}
}
\end{document}